\documentclass[twocolumn]{aastex631}
\hypersetup{linkcolor=blue,citecolor=blue,filecolor=cyan,urlcolor=blue}

\graphicspath{{figures/}{../figures/}}
\usepackage{newtxtext,newtxmath}
\usepackage[T1]{fontenc}

\usepackage{graphicx}	
\usepackage{amsmath}	
\usepackage{amssymb}	
\usepackage{hyperref} 
\usepackage{natbib}

\newcommand{\msun}{\ensuremath{\: \rm{M_{\odot}}}}
\newcommand{\popiii}{\ensuremath{\rm Pop~III}}
\newcommand{\popii}{\ensuremath{\rm Pop~II/I}}
\newcommand{\vl}{\ensuremath{``}}
\newcommand{\vd}{{''}}

\accepted{ApJ May, 2025}

\begin{document}

\title{Hidden Population III Descendants in Ultra-Faint Dwarf Galaxies}

\author[0000-0001-6887-2663]{Martina Rossi}
\affiliation{Dipartimento di Fisica e Astronomia “Augusto Righi”, Alma Mater Studiorum, Università di Bologna, Via Gobetti 93/2, 40129 Bologna, Italy}
\affiliation{INAF – Osservatorio di Astrofisica e Scienza dello Spazio di Bologna, Via Gobetti 93/3, 40129 Bologna, Italy}
\affiliation{Dipartimento di Fisica e Astrofisica
Univerisitá degli Studi di Firenze,
via G. Sansone 1, Sesto Fiorentino, Italy}

\author[0000-0001-7298-2478]{Stefania Salvadori}
\affiliation{Dipartimento di Fisica e Astrofisica
Univerisitá degli Studi di Firenze,
via G. Sansone 1, Sesto Fiorentino, Italy}
\affiliation{INAF/Osservatorio Astrofisico di Arcetri, Largo E. Fermi 5, I-50125 Firenze, Italy}

\author[0000-0001-9155-9018]{{\'A}sa Sk{\'u}lad{\'o}ttir}
\affiliation{Dipartimento di Fisica e Astrofisica
Univerisitá degli Studi di Firenze,
via G. Sansone 1, Sesto Fiorentino, Italy}

\author[0000-0001-9647-0493]{Irene Vanni}
\affiliation{Dipartimento di Fisica e Astrofisica
Univerisitá degli Studi di Firenze,
via G. Sansone 1, Sesto Fiorentino, Italy}
\affiliation{INAF/Osservatorio Astrofisico di Arcetri, Largo E. Fermi 5, I-50125 Firenze, Italy}

\author[0000-0002-3524-7172]{Ioanna Koutsouridou}
\affiliation{Dipartimento di Fisica e Astrofisica
Univerisitá degli Studi di Firenze,
via G. Sansone 1, Sesto Fiorentino, Italy}



\begin{abstract}

The elusive properties of the first (Pop~III) stars can be indirectly unveiled by uncovering their true descendants. To this aim, we exploit our data-calibrated model for the best-studied ultra-faint dwarf (UFD) galaxy, Boötes~I, which tracks the chemical evolution (from carbon to zinc) of individual stars from their formation to the present day. We explore the chemical imprint of Pop~III supernovae (SNe), with different explosion energies and masses, showing that they leave distinct chemical signatures in their descendants.
We find that UFDs are strongly affected by SNe-driven feedback resulting in a very low fraction of metals retained by their gravitational potential well ($<2.5\%$). Furthermore, the higher the Pop~III SN explosion energy, the lower the fraction of metals retained. 
Thus, the probability to find descendants of energetic  Pair Instability SNe is extremely low in these systems. Conversely, UFDs are ideal cosmic laboratories to identify the fingerprints of less massive and energetic Pop~III SNe through their [X/Fe] abundance ratios. Digging into the literature data of Boötes~I, we uncover three hidden candidates for Pop~III descendants: one mono-enriched and two multi-enriched. These stars show the chemical signature of Pop~III SNe in the mass range $[20-65] \msun$, spanning a wide range in explosion energies $[0.3- 10]10^{51}$~erg. In conclusion, candidates for Pop~III descendants are hidden in ancient UFDs but those mono-enriched by a single Pop~III SN are extremely rare. Thus, self-consistent models such as the one presented here are required to uncover these precious fossils and probe the properties of the first Pop~III supernovae.

\end{abstract}


\section{Introduction} \label{sec:intro}
A few hundred million years after the Big Bang, in the middle of the Dark Ages, the first (Pop~III) stars lit up the Universe initiating 
its radical transformation. Pop~III stars produced the first ionizing photons and while evolving as supernovae (SNe) they injected 
the first heavy elements, i.e., metals, into the surrounding gas. Understanding the properties of these first cosmic sources, such as their  
Initial Mass Function (IMF) and the explosion energy of Pop~III SNe, is thus crucial to study the early phases of reionization 
and metal enrichment.\\ 

\noindent The first stars are predicted to form at redshift $z \sim 20-30$ in {\it{minihaloes}} of primordial composition gas and thus to be 
metal-free \citep[e.g.][]{abel02, Bromm13}. Owing to the lack of efficient coolants in their birth environment, which implies more massive 
proto-stellar gas clouds \citep[e.g.][]{bromm01, Omukai05} and higher gas accretion rate onto the proto-star \citep{omukai01}, we expect Pop~III 
stars to be more massive than present-day stars, reaching masses up to $1000 \: \rm M_{\odot}$ \citep[e.g.][]{tan04, susa14, Hirano14}. 
Nevertheless, some 3D hydrodynamical simulations show that zero-metallicity gas clouds can fragment in sub-solar clumps \citep[e.g.][]{greif11, Prole22, Jaura+22},
possibly leading to the formation of clusters of low-mass long-lived Pop~III stars, which can survive until the present day - but see also \cite{sharda21} for different findings. 
However, these 3D simulations can only follow the first star formation for a very short time ($\approx 10^3$yr). 
Thus, on longer time-scales, the frequent mergers experienced by these clumps \cite[e.g.][]{HiranoBromm2017}, combined with the high accretion 
rate of the pristine gas onto the proto-stars, likely led to the formation of massive and very massive Pop~III stars \citep[e.g., see][for a recent review]{Klessen23}.\\ 

\noindent From an observational prospective, the idea that Pop~III stars are more massive than present-day stars is supported by the persistent lack 
of Pop~III survivors. Despite intensive searches, no metal-free stars have been observed in the Local Universe, suggesting that if low-mass long-lived 
Pop~III stars ($m_{\star}<0.8 \msun$) exist they must be extremely rare \citep[e.g.][]{SS07, Hartwig15, Magg19, Rossi+21}. 
In conclusion, the Pop~III IMF was almost certainly different than the present-day one and it was possibly biased towards more massive stars, with a 
characteristic mass, $m_{ch}>1 \msun$ \citep[e.g.][]{debennessuti17, sarmento19, Rossi+21, pagnini+23}. \\ %

\noindent If Pop~III stars are predominantly massive, then most of them are expected to evolve in a few Myr exploding as SNe and polluting the surrounding interstellar medium (ISM) with their newly produced chemical elements. In the mass range $m_{\star}=[10- 100]~\rm M_{\odot}$, each 
Pop~III SNe can evolve with very different explosion energy, from $\rm E_{SN} \approx 10^{50}$~erg to $\rm E_{SN} \approx 10^{52}$~erg (e.g. \citealt{Heger+woosley10, kobayashi06}). Conversely, very massive Pop~III stars, $m_{\star}=[140-260]\msun$, explode as powerful Pair Instability Supernovae (PISN), with an energy that increases with their progenitor mass \citep{heger02, Takahashi2018}. The amount and composition of new elements produced and injected in the ISM by these first Pop~III SNe depend upon both the progenitor mass and the SNe explosion energy (see e.g. \citealt{Vanni23}).\\

\noindent The chemical signatures of these pristine SNe can be retained in the photospheres of ancient, low-mass Pop~II stars that formed from \vl the ashes\vd\ of these massive first stars.\\

\noindent In the Local Group, the pursuit of identifying the descendants of Pop~III  stars has yielded fruitful results, enabled by the capacity to observe and resolve individual stars.   
Among ancient very metal-poor stars ($\rm [Fe/H]<-2$), the most promising candidates are the so called Carbon-Enhanced Metal Poor - no (CEMP-no) stars, which are characterized by [C/Fe] > +0.7 \citep[e.g.][]{beers05,bonifacio15} and no excess in neutron-capture elements, $\rm [Ba/Fe] < 0$. Various studies have shown a connection between CEMP-no stars and the chemical elements produced by Pop~III stars \citep{Iwamoto05, ishigaki14}, in particular those that explode as {\it faint supernovae} ($\rm E_{\rm SN} < 10^{51} \: erg$). On the other hand, the Pop~III  descendants imprinted by \textit{energetic} pristine SNe ($\rm E_{SN}>10^{51} \: \rm erg$) have been found both among CEMP-no stars and Carbon-normal, $\rm[C/Fe]<+0.7$, stars:
\cite{Ezzeddine2019} identified the fingerprints of asymmetric {\it{hypernova}} with $ \rm E_{SN} = 5 \times 10^{51} \: erg$ in the CEMP-no star HE 1327-2326 at $\rm [Fe/H] \approx -5$ and more recently, \cite{Skuladottir2021} and \cite{Placco21} discovered two C-normal metal-poor stars ($\rm [Fe/H]< -4$) in the Sculptor dwarf galaxy and in the Galactic halo, respectively, whose abundance patterns are compatible with an enrichment by a Pop~III {\it{hypernova}}, with $ \rm E_{SN} = 10 \times 10^{51} \: erg$.\\

\noindent Despite these discoveries at extremely low-metallicities, models predict that the probability to find 
direct descendants of high-energy Pop~III SNe increases towards higher [Fe/H] (\citealt{Vanni23}). However, their detection becomes more challenging at higher [Fe/H] due to the increasing numbers of subsequent generations of Pop~II descendants there \citep{SS15, debennessuti17, Hartwig+18, ioanna2023}. 
For example, the descendants of PISN are expected to appear in a very broad metallicity range $\rm -4 < [Fe/H] < -1$ with a peak at $\rm [Fe/H] \approx -1.8$ \citep{karlsson13,  debennessuti17, SS19, ioanna2024}. Indeed, several
candidate PISN descendants have been found at [Fe/H]$\approx -2$ \citep{aguado23} including the one identified by \cite[][]{xing23}, whose abundance pattern is unfortunately not anymore consistent with a pure PISN enrichment (\citealt{Asa24PISN, Thibodeaux24}).\\

In conclusion, the precious \popiii\ descendants are expected to appear in a wide range of metallicity from $\rm [Fe/H] \approx -1$  down to $\rm [Fe/H] \approx -7$ and among both CEMP and C-normal stars.\\

Among all environments hosting ancient stars, Ultra-Faint Dwarf galaxies (UFDs) stand out as ideal systems for studying  the nature of the first stars and catching the chemical finger-prints of the first SNe \citep{SS09, SS15, magg17, Hartwig+19, Rossi+21, Rossi23}. From a theoretical perspective, these low-mass dwarf galaxies are compelling candidates for being the building blocks of present-day galaxies, such as our Milky Way, and the first star forming systems hosting Pop~III stars (\citealt{SS15}). Observationally, they are the oldest, most dark matter-dominated, most metal-poor, least luminous ($\rm L_{\rm bol} < 10^5 \: L_{\odot}$), and least chemically evolved stellar systems known (\citealt[]{Simon19}). The majority of UFDs formed more than 75\% of their stars within the first billion years of their evolution, resulting in truly ancient stellar populations, with ages exceeding 12 Gyr \citep{Brown14,Gallart+21}. Furthermore, UFDs harbor the highest fraction of very metal-poor stars ($\rm [Fe/H]<-2$), with a significant portion of CEMP-no stars \citep{kirby13, yong+13, spite2018, Yoon19}. Finally, the observed stars in UFDs display a broad metallicity range of  $\rm -4 < [Fe/H] < -1$ (\citealt{Fu2023}). These properties make UFDs compelling laboratories to search for the chemical imprints of Pop~III SNe with different energies, ranging form \textit{faint} ($\rm E_{SN} \approx 10^{50} \: erg$) SNe up to PISN ($\rm E_{SN} > 10^{52} \: erg$).
\\

\noindent In this work, we investigate the chemical signatures of Pop~III SNe with different energies and masses, in UFD galaxies especially focusing on one of the most luminous and best studied, Boötes~I.
\section{Model description}

\label{model}

The data-calibrated model used in this work was first introduced in \cite{Rossi+21} and expanded in \cite{Rossi23}. The aim of the model is to
describe the chemical evolution of an UFD, particularly focusing on the best studied system: Boötes~I. 
This semi-analytical model follows the star formation and chemical enrichment history of Boötes~I from its formation epoch to the present day, by tracking elements from carbon to zinc produced 
by both Pop~III and Pop~II/I stars. The model can be summarized as follows (for details see \citealt{Rossi+21, Rossi23}):\\

\begin{figure}
    \centering
    \includegraphics[width=\columnwidth]{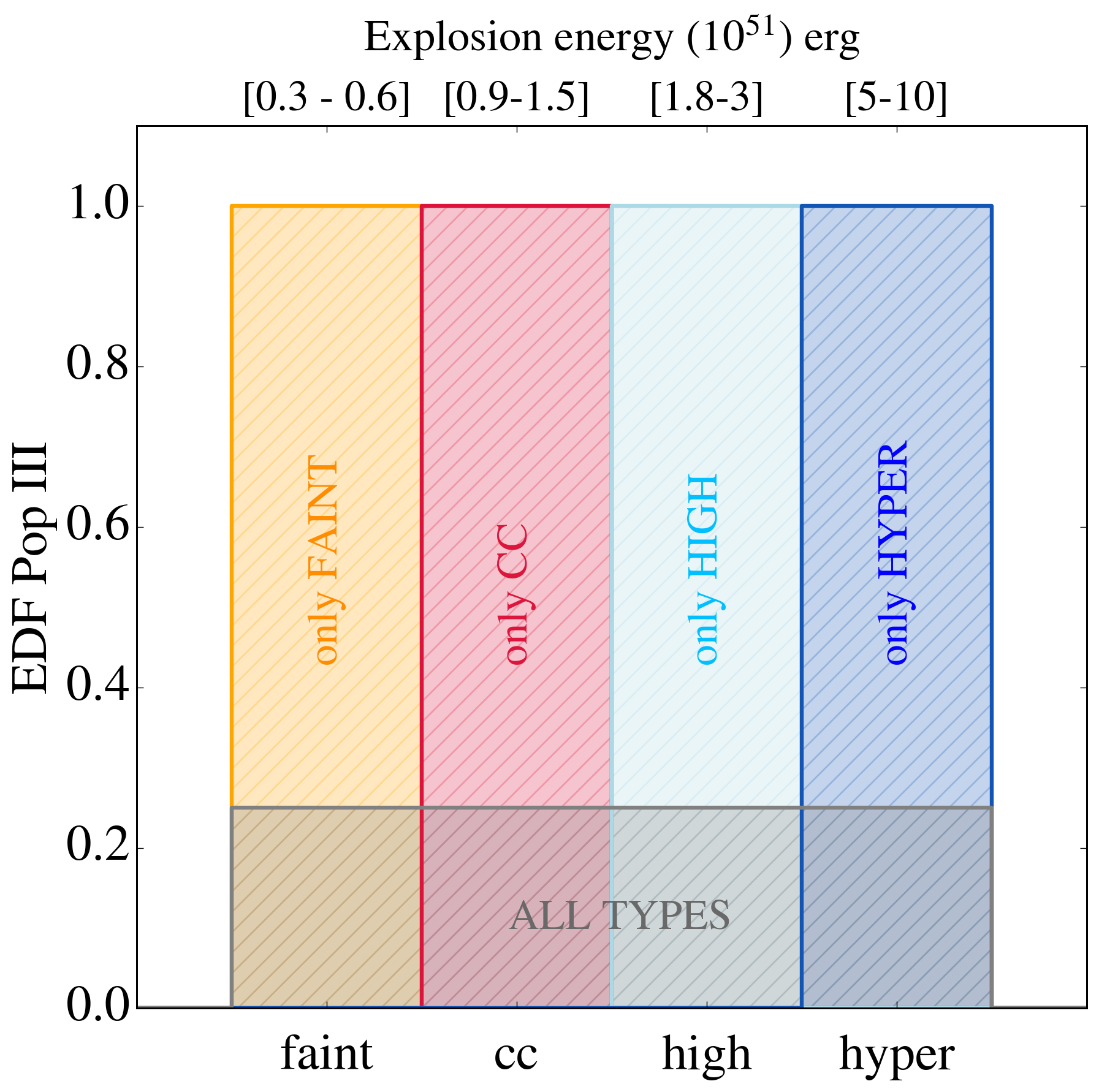}
    \vspace{0.5cm} 
    \includegraphics[width=\columnwidth]{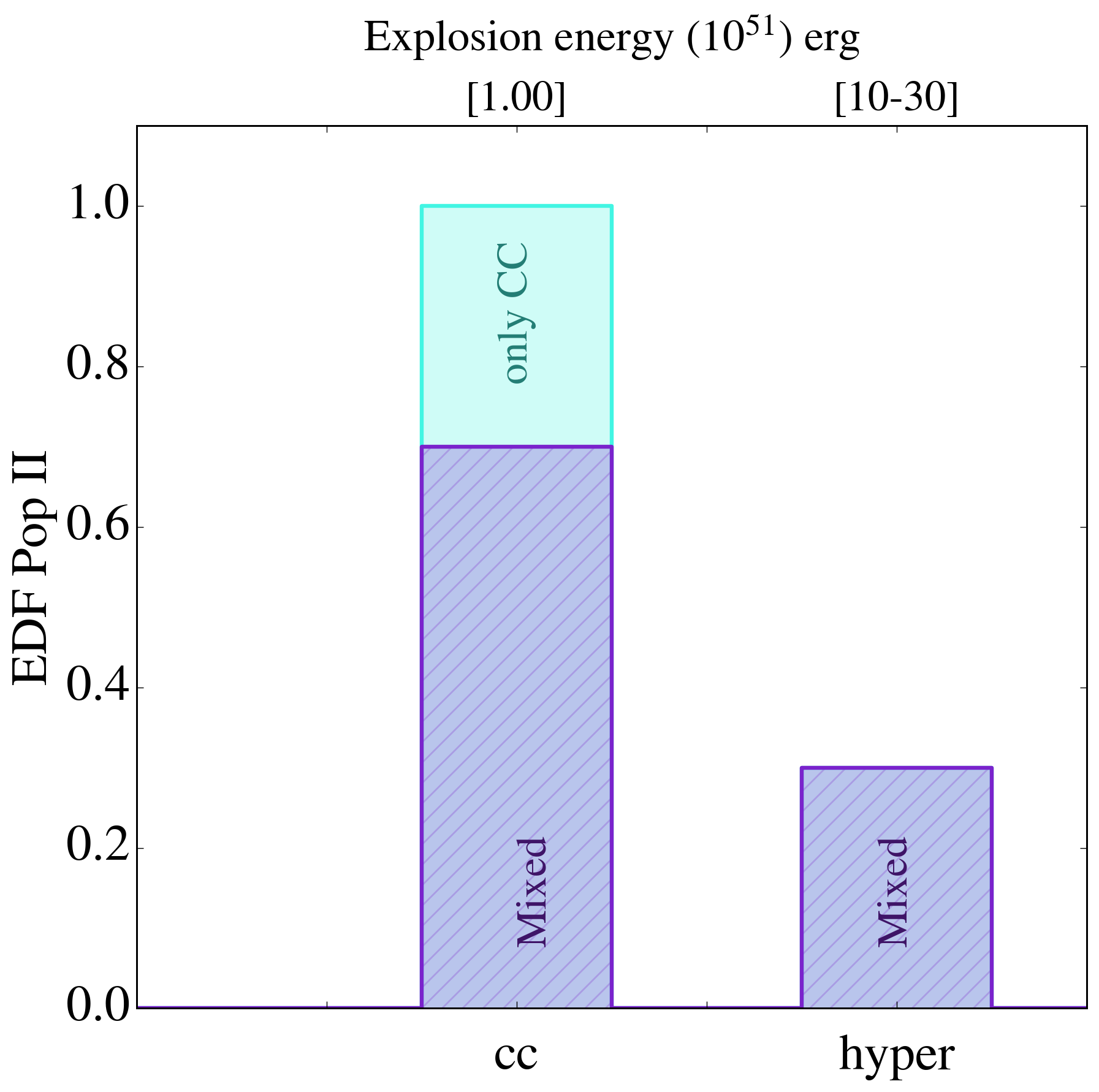}
    \caption{Assumed EDFs for Pop~III (top panel): four cases in which the EDF is a top-hat function centered on each type of SNe ($100\% $ probability), and the last case in which the all SNe explosion energies are equiprobable ($25\%$ probability of each type).  {Assumed EDF for Pop~II/I stars (bottom panel): the first case assumes a top-hat EDF centered on the explosion energy of core-collapse (CC) supernovae (100\% probability); while the second case considers a mixed distribution where $70\%$ of Pop~II/I SNe have CC energies, and $30\%$ correspond to hypernovae.} }
    \label{edf_teo}
\end{figure}

{\bf Initial conditions:} We assume Boötes~I to evolve in isolation, i.e. without experiencing major merger events.
This is consistent with cosmological models \citep{SS09, SS15}, which are used to set 
the initial conditions: the epoch of virialization, $\rm z_{vir}\approx 8.5$,  {and the corresponding dark matter halo mass at that redshift}, $\rm M_{h}\approx 10^{7.3} \msun$.\\

{\bf{Star formation:}} At the virialization epoch we compute the total amount of gas\footnote{We adopt $\Lambda$ cold dark matter ($\Lambda$CDM) with $h = 0.669$, $\Omega_{\rm m}=0.3103$, $\Omega_{\Lambda}=0.6897$, $\Omega_{\rm b} h^2 = 0.02234$, $\sigma_8 = 0.8083 $, $n= 0.9671$, according to the latest Plank results \citep{plank18}} as $M_g = \Omega_b/\Omega_m \: \rm M_{h}$, and we assume it to be pristine. The star-formation becomes possible when the gas begins to fall into the central part of the halo and cool down \citep[see][for the infall rate]{Rossi+21}. We evaluate the total mass of stars formed in a timescale of 1~Myr, assuming a 
star-formation rate regulated by  {the redshift dependent} free-fall time ($\rm t_{ff}(z)$) and the  mass of cold gas accumulated via infall ($\rm M_{cold}$): $\rm \Psi = \epsilon_\star M_{cold}/t_{ff}(z)$, where $\epsilon_\star$ is a free parameter of the model, equal for both Pop~III and Pop~II/I stars.  {Note that  since the free-fall time  $\rm t_{\text{ff}}(z)$  is larger at high redshift (i.e., when Pop~III stars form), this results in a smaller star formation rate ($\Psi$) for Pop~III stars compared to Pop~II/I stars, even if  $\epsilon_\star$  is the same. }\\

Following the critical metallicity scenario \citep{bromm01, Schneider2003, Omukai05}, Pop~III stars only form if the metallicity of the ISM gas\footnote{ $\rm Z_{gas} = M^{ISM}_{Z}/M_{gas}$ where $\rm M^{ISM}_{Z}$ and $\rm M_{gas}$ are the mass of metals and gas in the ISM, respectively.}, $\rm Z_{gas}$, is below the critical metallicity, $\rm Z_{cr} = 10^{-4.5 \pm 1}Z_{\odot}$, i.e. $\rm Z_{gas} \leq Z_{cr}$, while {\it{normal}} Pop~II stars form when  $\rm Z_{gas} > Z_{cr}$. \\

{\bf{Initial Mass Function (IMF):}} We assume the newly-formed stellar mass to be distributed according to an Larson IMF \citep{Lars98}:
    \begin{equation}
    \label{imf}
        \phi(m_{\star}) = \frac{dN}{dm_{\star}} \propto m_{\star}^{-2.35} exp(-m_{ch}/m_{\star}) 
    \end{equation}
where $m_{\star}$ is the stellar mass and $m_{ch}$ the characteristic mass of the IMF. For Pop~III stars $m_{ch }= 10 \: \rm M_{\odot}$ and $m_{\star}=[0.8-1000]\msun$ (consistent with the results of \citealp{Hirano14, HiranoBromm2017, Rossi+21, pagnini+23}); while for Pop~II/I stars $m_{ch}= 0.35 \: \rm M_{\odot}$ and $m_{\star}=[0.1-100]\msun$.\\

\noindent The model includes the incomplete sampling of the stellar IMF for both Pop~III and Pop~II/I stellar populations. Every time that the conditions for star formation are reached, a discrete number of stars is formed, with mass selected randomly from the assumed IMF. This is a key feature of the model, essential for accurately simulating the evolution of poorly star-forming UFDs in which the assumed IMF is never fully sampled  throughout its entire evolution (see \citealt{Rossi+21}). \\


{\bf{Chemical feedback:}} The chemical enrichment of the gas is followed by taking into account the mass-dependent timescales for the evolution of individual stars.  {Stellar lifetimes are calculated using the relation from \cite{Raiteri96}, which accounts for both stellar mass and metallicity, reflecting the chemical composition of the ISM from which the stars originated.}
    The semi-analytical model tracks and follows the chemical elements (from carbon to zinc) released by SNe and AGB stars from both the \popiii \: and \popii \: stellar populations.
    
    \begin{itemize}
        \item SNe can enrich the ISM with all chemical elements from carbon to zinc.
    For \popiii \: SNe in the mass range $m_{\popiii}= [10-100] \msun$ we use the stellar yields from \cite{Heger+woosley10}, while for massive PISN we adopt the ones of \cite{heger02} in the mass range $m_\text{PISN}= [140-260] \msun$. For \popii \: SNe in the mass range $[8-40]\msun$ we use  {three set of stellar yields:  \cite{Limongi+18} (hereafter LC), specifically model R, which assumes no rotation and a fixed explosion energy of $\rm E_{SN} = 10^{51} erg$; the yields from \cite{NK13}, including only core-collapse Pop~II SNe with $\rm E_{SN} = 10^{51}  erg$ (hereafter NK CC); and \cite{NK13} models with hypernove (hereafter NK CC+HN) with $\rm E_{SN} = [10-30] \times 10^{51} , erg$ depending on the stellar mass in the range [13-40]$\msun$.}
    \item  AGB stars  in the mass range $[2-8]\msun$ enrich the ISM through stellar winds, only in carbon, nitrogen, and oxygen. We adopt the stellar yields of \cite{Meynet+02} (rotating model) for \popiii \: stars, and \cite{VanDenHoek+97} for \popii \: stars.
    \item  {Type Ia supernovae (SNIa): The contribution of SNIa has been included by adopting the yields by \cite{Iwamoto99} and the bimodal delay time distribution observationally derived by \cite{Mannucci+06}. At each time-step and for each stellar burst we computed the rate of SNIa by following \cite{matteucci06}, where the normalization constant has been fixed to reproduce the actual rate of SNIa in the Milky Way.}
    \end{itemize}
    
{\bf{Mechanical feedback:}} Stars that end their lives as SNe not only pollute the ISM with heavy elements but can also enrich the intergalactic medium (IGM) through mechanical feedback. Indeed, SNe explosions can generate a blast wave that, when sufficiently energetic, can overcome the gravitational potential well of the host halo, resulting in the expulsion of gas and metals into the IGM.
The ejected mass, $\rm M_{ej}$ is regulated by: 
$\rm M_{ej} = (2 \epsilon_{w} {N_{SN}} E_{SN})/{v^{2}_{esc}}$ \citep[see][]{SS08}. Here, $\epsilon_{w} \rm N_{SN} \rm E_{SN}$ is the total kinetic energy injected into the halo, $\rm N_{SN}$ the number of SNe, $\rm E_{SN}$ the explosion energy, 
and $\epsilon_{w}$ is a free parameter, which controls efficiency of the SNe-driven winds. The escape velocity of the gas depends on the halo mass, $M_h$, and viral radius of the halo \citep[see][]{Bark01}). The metals 
injected into the ISM by SNe undergo complete mixing with the halo gas, so the metallicity of the ejected gas is the same as that of the ISM.\\

{\bf{Radiative feedback:}}  {Our model does not explicitly incorporate radiative feedback processes; however, the halo mass selected for the Boötes~I dwarf galaxy is derived from cosmological models that account for the increase in the minimum mass required for star formation due to the influence of the Lyman-Werner background. The model do not account for internal radiative feedback from young stars which may have a role in reducing the efficiency of star formation of low-mass galaxies (\cite{Wise12, jeon+14}).}\\

{\bf{Energy Distribution Function (EDF) of Pop~III SNe:}} For $[10-100] \msun$ Pop~III stars ending their lives as SNe, we follow \cite{Heger+woosley10} and explore different possible explosion energies :

    \begin{itemize}
        \item \emph{faint} SNe: $ \rm E_{SN}=[0.3-0.6] \times 10^{51}$~erg;
        \item \emph{core-collapse (cc)} SNe: $\rm E_{SN}=[1.2-1.5] \times 10^{51}$~erg;
        \item \emph{high-energy} SNe: $\rm E_{SN}=[1.8-3.0] \times 10^{51}$~erg;
        \item \emph{hypernovae}:  $\rm E_{SN}=[5.0-10.0] \times 10^{51}$~erg;
    \end{itemize}

We explore five different possibilities for the unknown EDF of Pop~III SNe, shown in Fig.\ref{edf_teo}  {(top panel)}. In the first four cases we assume extreme scenarios in which Pop~III SNe are all of a single type: {\it{only faint}}, {\it{only cc}}, {\it{only high}}, or {\it{only hyper}}. This corresponds to a top-hat EDF in which the probability to form the chosen type of SNe is $100\%$. In the {fifth} case we assume an EDF spread among all different explosion energies (\vl all types\vd\ model), where each SNe type has an equal probability of $25\%$. Thus, every time a Pop~III SN progenitor forms we randomly assign to it an explosion energy. 

\noindent Conversely, in the mass range $[140-260]\msun$, we assume that \popiii \: stars evolve as PISNe with an explosion energy $\rm E_{SN}=[10, 100]\times 10^{51} erg$, proportional to the stellar mass (\citealt{heger02}), and therefore {\it independent} from the assumed EDF for SNe with less massive progenitor stars. Note that the probability to form PISNe is regulated by both the assumed Pop~III IMF and the star-formation rate, which affect the random sampling of the Pop~III IMF \citep[see][]{Rossi+21}. Assuming a fully sampled Pop~III IMF (with $m_{ch} = 10 \msun$) we obtain that this probability is $\sim 0.2\%$.  \\

{\bf{EDF of Pop~II/I SNe:}}
 {For Pop~II/I stars that end their lives as SNe, we explore the two EDFs shown in Fig.\ref{edf_teo} (bottom panel)}:
\begin{itemize}
\item {\textit{core-collapse}}:  {A flat EDF where all Pop~II/I SNe have a fixed explosion energy of $\rm E_{SN} = 10^{51}$~erg. The stellar yields associated to this EDF are derived from either \citet{Limongi+18} or \citet{NK13}. In the LC model, the stellar SNe mass range is $[8-40] , \msun$, while for the NK CC model, it is $[13-40] , \msun$.}
\item {\textit{core-collapse + hypernovae}}: { We also consider an EDF where $70\%$ of Pop~II/I SNe are core-collapse supernovae with a fixed explosion energy of $\rm E_{SN} = 10^{51}$~erg, while the remaining $30\%$ are hypernovae \cite{ishigaki14}. The hypernovae explosion energy is mass-dependent and spans the range $\rm E_{SN} = [10-30] \times 10^{51}$~erg.}

\end{itemize}

\begin{table*}[!ht]
    \centering
    \begin{tabular}{|c|c|c|c|c|c|}
    \hline
         &  $\rm \log(L_{\star}/L_{\odot}) $ & $t_{50} \rm (Myr)$& $\rm \chi^2_{MDF}$ &$(\epsilon_{\star}, \epsilon_{w})$ & $\chi^2_{\nu}$  \\ \hline
         {LC CC yields} & ~ & ~ & ~ & ~ & ~ \\ \hline
        FAINT & $4.69 \pm 0.03$ & $24 \pm 1$ & 186 &(1.0, 0.009) & 25.92 \\ 
        CC & $4.78 \pm 0.05$ & $29 \pm 3$ & 40 &(1.0, 0.008) & 12.57 \\ 
        HIGH & $4.61 \pm 0.03$ & $33 \pm 1$ & 23 &(0.5, 0.0014) & 5.71 \\ 
        HYPER & $4.58 \pm 0.26$ & $37 \pm 2$ & 37 &(0.5, 0.001) & 6.20 \\ 
        ALL TYPE & $4.62 \pm 0.06$ & $33 \pm 2$ & 18 &(0.9, 0.002) & 5.05 \\ 
 \hline \hline
         {NK CC yields} & ~ & ~ & ~ & ~ & ~ \\ \hline
        FAINT & $5.29 \pm 0.01$ & $33 \pm 1$ & 80 &(0.5, 0.002) & 71.54 \\ 
        CC & $5.32 \pm 0.05$ & $34 \pm 1$ & 21 &(1.2, 0.001) & 56.16 \\ 
        HIGH & $5.27 \pm 0.11$ & $34 \pm 1$ & 30 &(1.2, 0.001) & 31.05 \\ 
        HYPER & $4.34 \pm 0.35$ & $37 \pm 2$ & 89 &(0.5, 0.001) & 12.17 \\ 
        ALL TYPE & $4.97 \pm 0.04$ & $35 \pm 1$ & 24 &(0.4, 0.002) & 23.47 \\ 
        
 \hline \hline
         {NK CC+HN yields} & ~ & ~ & ~ & ~ & ~ \\ \hline
        FAINT & $5.24 \pm 0.02$ & $31 \pm 1$ & 77 &(1.2, 0.002) & 63.32 \\ 
        CC & $5.28 \pm 0.05$ & $33 \pm 1$ & 48 &(1.1, 0.001) & 56.27 \\ 
        HIGH & $5.24 \pm 0.10$ & $33 \pm 1$ & 26 &(1.2, 0.001) & 33.30 \\ 
        HYPER & $4.89 \pm 0.10$ & $34 \pm 1$ & 42 &(0.5, 0.004) & 14.03 \\ 
        ALL TYPE & $5.06 \pm 0.07$ & $33 \pm 1$ & 14 &(0.6, 0.001) & 25.00 \\ 
 \hline \hline
    \end{tabular}
    \caption{Simulated properties of Bo\"otes~I with corresponding $\chi^2$ values, using using Pop~III yields from \citet{Heger+woosley10}; and different Pop~I/II yields as listed. The EDFs are adopted from Fig.~\ref{edf_teo}). The observed properties of Bo\"otes~I include the total luminosity, $\log(L_{\star}/L_{\odot})=4.5\pm0.1$ \citep{kirby13}, the time needed to form $50\%$ of the total stellar mass ($t_{50}$, \citealt{Brown14}), and the MDF \citep{Jenkins2021}, see text for details.
    }
    \label{free_para}
\end{table*}

 {
\section{Model calibration}}
{The model has been calibrated using observational data as detailed in \citep{Rossi+21}. This means that the two free parameters ($\epsilon_{\star}$, $\epsilon_{w}$) are fixed to match the observed properties of Boötes~I: the total luminosity, $\rm \log(L_{\rm obs}/ L_{\odot}) = 4.5 \pm 0.1$ \citep{kirby13}; the Metallicity Distribution Function (MDF) (\cite{Jenkins2021},  which is known to be one of the key observables to constrain chemical evolution models \citep[e.g.,][]{salvadori+07,koutsouridou+23}; and the time needed to form 50\% of the total stellar mass, $t_{50}$, which provides the weakest constraints ranging from less than $100$~Myr to $\approx 1$~Gyr \citep{Brown14}. Compared to \cite{Rossi+21}, we have updated the Boötes~I stellar sample using data from \cite{Jenkins2021}, consisting of 54 stars. This results in a revised mean metallicity of $\langle \rm [Fe/H] \rangle = -2.34 \pm 0.28$}.

Due to the stochastic nature of the IMF sampling, the chemical enrichment history of Boötes~I varies across different runs of the simulation. To account for this variability, all results presented here are derived by averaging over 1000 independent realizations of the galaxy’s evolutionary history. The associated scatter across these runs is quantified as the standard deviation.

 {From the ensemble of realizations, we compute the mean simulated values ($\langle \rm X_{\rm sim} \rangle$) and their associated uncertainties ($\sigma_{\rm sim}$) for the observables mentioned above (luminosity, MDF, and $t_{50}$). These values are used to compute the total $\chi^2$ as the sum of three terms, each corresponding to a specific observable:}
\begin{equation}
\chi^2 = \chi^2_{\rm Lum} + \chi^2_{t_{50}} + \chi^2_{\rm MDF}.
\end{equation}

 {The terms $\chi^2_{\rm Lum}$ and $\chi^2_{t_{50}}$ are computed using the general formula:}
\begin{equation}
\chi^2_{\rm X} = \frac{\left(X_{\rm obs} - \langle X_{\rm sim} \rangle \right)^2}{\sigma_{\rm X, obs}^2 + \sigma_{\rm X, sim}^2},
\end{equation}
 {where $X$ represents the observable: total luminosity (${\rm Lum}$) or $t_{50}$. The observed values ($X_{\rm obs}$) are compared to the mean simulated values ($\langle X_{\rm sim} \rangle$), and the combined uncertainties from observations ($\sigma_{\rm X, obs}$) and simulations ($\sigma_{\rm X, sim}$) are used in the denominator.}

 {The third contribution to the total $\chi^2$ is derived from the MDF. For each run of the simulation, we compute the MDF by binning stars in intervals of [Fe/H]. The mean simulated MDF ($\langle {\rm MDF}{\rm sim} \rangle$) is then calculated by averaging the number of stars in each iron bin across 1000 independent realizations, and the associated uncertainty  is quantified as the standard deviation among these runs. Similarly, the observational MDF is characterized by the observed number of stars in each [Fe/H] bin ($N{\rm obs}$), with the uncertainty on $N_{\rm obs}$ in each bin assumed to be $\sigma_{\rm MDF, obs} = \sqrt{N_{\rm obs}}$. The $\chi^2_{\rm MDF}$ term quantifies the agreement between the simulated and observed MDF and is computed as:}
\begin{equation}
\chi^2_{\rm MDF} = \sum_{k=1}^{n} \frac{\left(N_{\rm obs}[k] - \langle N_{\rm sim}[k] \rangle \right)^2}{\sigma_{\rm MDF, obs}[k]^2 + \sigma_{\rm MDF, sim}[k]^2},
\end{equation}
 {where $n$ is the number of [Fe/H] bins, $N_{\rm obs}[k]$ is the number of observed stars in the $k$-th bin, $\langle N_{\rm sim}[k] \rangle$ is the mean simulated number of stars in the same bin, and $\sigma_{\rm MDF, obs}[k]$ and $\sigma_{\rm MDF, sim}[k]$ are the uncertainties associated with the observed and simulated MDF, respectively. In this work, we include all bins with $\rm [Fe/H] > -4$.  Finally, we computed the reduced chi-square\footnote{ $\chi^2_{\nu} = \chi^2/ \nu$, where $\nu$ is the number of degrees of freedom, calculated as the total number of data points ($n_{\rm data}$) minus the number of free parameters in the model. In this work, $\nu = n_{\rm data} - 2 =9$, considering the two free parameters ($\epsilon_{\star}$, $\epsilon_{w}$) in the model }, $\chi^2_{\nu}$, to assess the goodness of fit of the model. Finally, we fix the free parameters ($\epsilon_{\star}$, $\epsilon_{w}$) by minimizing $\chi^2_{\nu}$.}
\\

\begin{figure*}
\centering
	\includegraphics[width=0.9\textwidth]{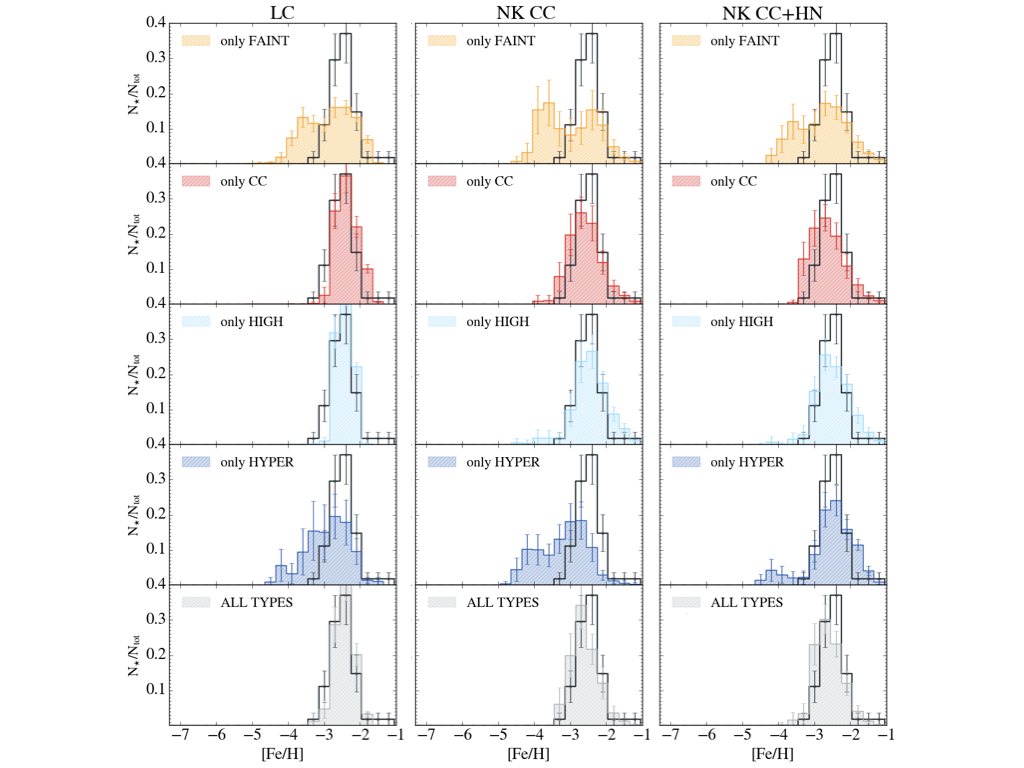}
    \caption{{Comparison of the observed (black outlines) and simulated (shaded colors) MDFs of Boötes~I. Columns show different assumptions for the Pop~II/I EDFs/yields, while rows show different Pop~III EDFs.} }
    \label{MDFs}
\end{figure*}

\section{Fiducial Model}
 {To thoroughly investigate the chemical evolution of Boötes~I, we analyzed various combinations of Pop~II/I stellar yields and Pop~III EDFs. For the Pop~II/I stars, we considered three distinct sets of stellar yields: LC, NK CC, and NK CC+HN (see Sec.\ref{model}). For Pop~III stars, firstly we assume that all SNe in the mass range \([10-100] \, \msun\) belong to a single, unique type (i.e. \vl \textit{only faint}\vd, \vl \textit{only cc}\vd, \vl \textit{only high}\vd, and \vl \textit{only hyper}\vd). Then, we relax this assumption and adopt a flat \vl all type\vd\ EDF, where each of the four supernova types contributes equally, with a probability of 25$\%$ (see Fig.\ref{edf_teo}).} \\ 
 {For each combination of  Pop~II stellar yields and EDFs, we calibrated the model by determining the pair of parameters  ($\epsilon_{\star}$, $\epsilon_{\text{w}}$)  that minimizes the reduced chi-squared ( $\chi^2_{\nu}$ ) between theoretical predictions and observational constraints. Table \ref{free_para} summarizes the results and  lists the simulated values of $\rm \log(L_{\star}/L_{\odot})$, $t_{50}$, the contribution to the $\chi^2$ from the MDF, $\chi^2_{\rm MDF}$, and the $\epsilon_{\star}$, $\epsilon_{w}$ values that minimize the reduced chi-square and $\chi^2_{\nu}$.}\\

 {Our results show that the bulk of the chemical enrichment in Boötes~I is dominated by Pop~II/I stars. However, Pop~III stars determine the initial metallicity flow and chemical enrichment conditions on which Pop~II/I stars subsequently form, and therefore have a significant impact on the overall shape of the total MDF. }

 Nonetheless, we observe that to achieve a good fit as the explosion energy of Pop~III stars increases, it becomes necessary to either (1) reduce the star formation efficiency ($\epsilon_{\star}$), thereby forming fewer Pop~III SNe, or (2) decrease the wind efficiency ($\epsilon_{\text{w}}$), allowing more chemical elements from the energetic Pop~III SNe to be retained within the system. 
However, a clear trend is not always evident due to the stochastic nature of early galaxy evolution, driven by the random sampling of the IMF and the specific masses of Pop~III stars that are formed.

It is worth noting that the observational estimates of $t_{50}$ have very large uncertainties \citep[e.g.][]{Brown14}, ranging from $<100$\,Myr to $\sim1$\,Gyr, and they therefore provide very limited constraints. Our predicted values are systematically short ($\sim 30$ Myr), however, the $t_{50}$ values can be adjusted by including an additional free parameter, i.e., the gas infall time, here equal to $t_{\rm inf}\approx t_{\rm ff}(z)/4 \approx 15$~Myr \citep{Rossi+21}. By increasing $t_{\rm inf}$ we can obtain a better agreement with the observed $t_{50}$ without changes in the total luminosity nor the MDF (see Ciabattini et al. in prep.).
\\

 {The MDF plays a key role in constraining the models, as it is highly sensitive to both the Pop~III EDF and the choice of Pop~II stellar yields/EDFs. In Fig.\ref{MDFs}, the observed MDF of Boötes I \citep{Jenkins2021} is compared with the predicted MDFs for various combinations of Pop~III EDFs and Pop~II stellar yield sets (LC, NK CC, NK CC+HN). Each panel corresponds to a specific combination. The contribution to the total chi-squared arising specifically from the MDF is reported in the column $\chi^2_{\rm MDF}$ of Table~\ref{free_para}.} \\
 {Different combinations of Pop~III EDFs and Pop~II/I yields produce MDFs that vary significantly in shape and width.  For instance, models assuming “\textit{only faint}” PopIII SNe produce MDFs spanning a broader range of [\rm Fe/H], shifted toward lower metallicities.} Indeed faint SNe produce large amounts of C and small of Fe, resulting in a long low-Fe tail in the MDF which is not observed in Boötes~I (first row in Fig.~\ref{MDFs}), and therefore the \vl \textit{only faint}\vd\ model is less favoured. On the other hand, assuming that all Pop~III SNe explode as hypernovae shifts the peak of the MDF towards lower [Fe/H] values. Indeed, due to the low potential well of Boötes~I, when all Pop~III explode as energetic hypernovae the majority of the gas and metals are evacuated out of the galaxy. As a result, retaining metals in the halo becomes challenging. In addition, due to the lack of available gas for star formation, the star formation rate of subsequent Pop~II is very low ($\Psi \approx 10^{-3} \rm M_{\odot} yr^{-1}$), leading to less efficient metal enrichment.
 {It is worth noting that the “\textit{only high}”  models, where all Pop~III SNe explode as high-energy events, produce MDFs that at first glance appear to match the observed MDF. However, these models perform worse in reproducing other key observables of Boötes~I and are therefore strongly disfavored overall.} 

\begin{figure*}
\centering
	\includegraphics[width=0.8\textwidth]{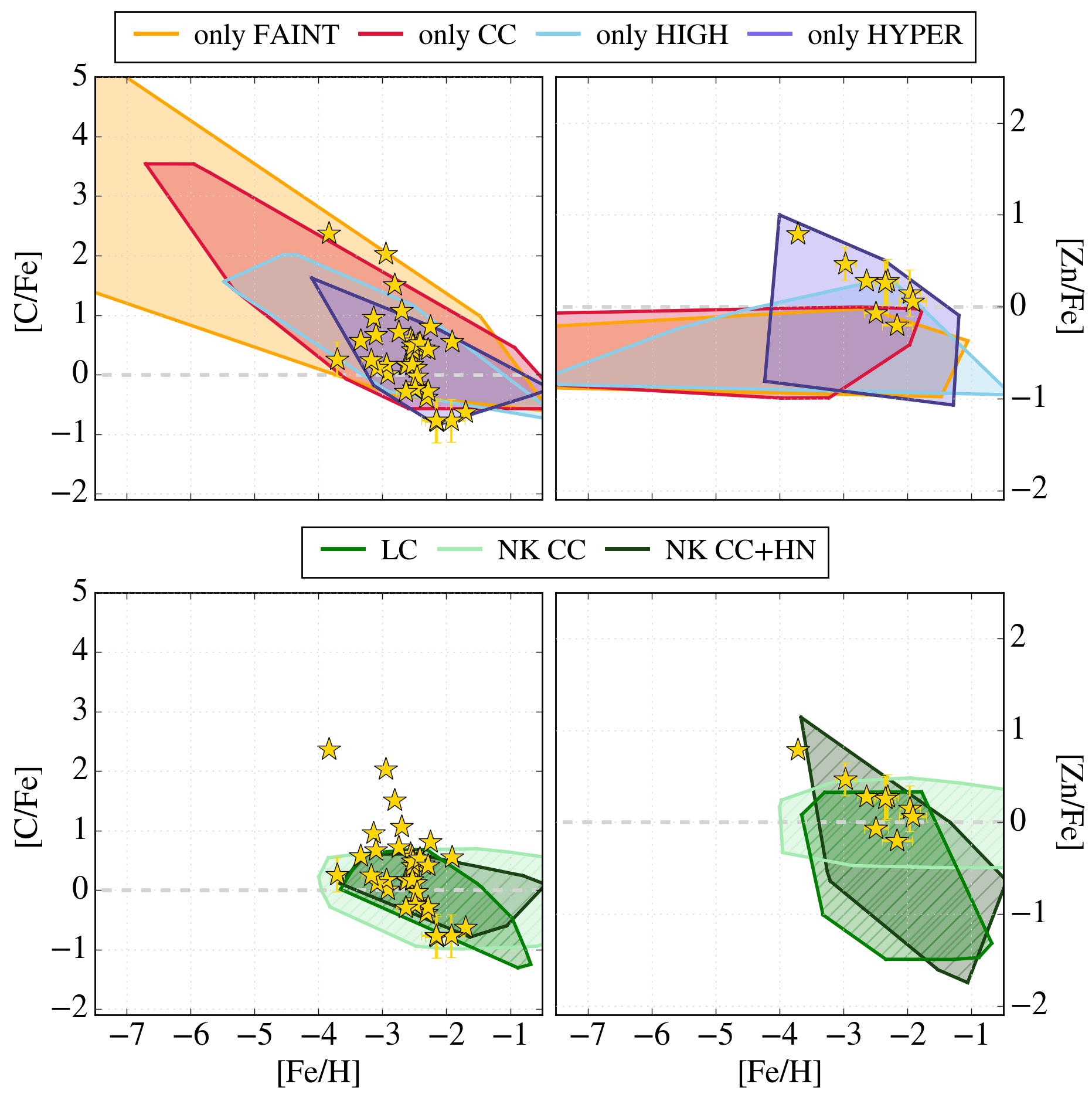}
    \caption{ The [C/Fe] (left) and [Zn/Fe] (right) abundance ratio of observed Boötes~I stars (yellow stars) as a function of their [Fe/H] (SAGA Database; \citealt{SAGA}). Top row shows the stellar populations enriched  by Pop~III SNe with different energies: {\it{faint}} (orange), {\it{cc}} (red), {\it{high}} (light blue), {\it{hyper}} (dark blue). {Bottom row shows the expected chemical enrichement by {\it{normal}} Pop~II/I SNe using different set of stellar yields (LC, NK CC, NK CC+HN)}. Note that the scales on the y-axis are different in the panels. 
    }
    \label{fig3}
\end{figure*}
 {The “all types” models, which assume an equal contribution from all Pop~III  types (faint, core-collapse, high-energy, and hypernovae), provides a more balanced distribution, successfully reproducing the intermediate metallicity range ($-3 \lesssim \rm [Fe/H] \lesssim -2$) observed in Boötes~I, independently of the Pop~II stellar yields adopted.} \\

 {The choice of Pop~II/I stellar yields has an even more pronounced effect, underscoring the dominant role of Pop~II/I stars in driving the chemical evolution of Boötes~I. For example, the LC yield set systematically produces lower $\chi^2$ values compared to NK CC and NK CC+HN, indicating a better agreement with the observed MDF and other constraints. This behavior can be traced back to the higher metal yields — particularly of iron — predicted by the NK yield set compared to LC. In order to reproduce the shape and normalization of the observed MDF, which carries the highest statistical weight in the total $\chi^2$ evaluation, the models compensate by adjusting the star formation efficiency ($\epsilon_\star$) and the outflow efficiency ($\epsilon_w$). Specifically, the enhanced Fe production per SN event in NK models generally requires a lower $\epsilon_\star$ to avoid an overproduction of metals. However, this leads to reduced stellar mass growth and lower predicted star counts, often resulting in tension with other constraints, such as the observed stellar luminosity. The overall worse performance of the NK-based models is therefore a consequence of the interplay between their nucleosynthetic yields and the global galactic parameters required to reproduce the MDF.
 Among the tested combinations, the “all types” Pop~III EDF paired with the LC Pop~II/I yields emerges as the best-fit model, achieving the lowest reduced $\chi^2$ value and providing a realistic reproduction of the observed properties of Boötes~I.} 

In Fig.~\ref{fig3} we analyze how the different EDFs for Pop~III SNe affect the predicted abundance ratios\footnote{$\rm [X/Fe]= \log(N_{X}/N_{Fe})- \log(N_{X, \odot}/N_{Fe, \odot})$; where $\rm N_{X}$ and $\rm N_{Fe}$ are the abundances of elements X and Fe, and $\rm N_{X, \odot}$ and $\rm N_{Fe, \odot}$ are the solar values \citep{Asplund+09}} of present-day stars in Boötes~I. In particular we focus on [C/Fe] and [Zn/Fe], which are very sensitive to different explosion energies.
For each model, the contours delineate the stellar populations that have been {\it predominantly imprinted} by a unique type of Pop~III SNe, i.e., that contributed $> 95\%$ of their present-day mass in metals. 
The location of these Pop~III descendants in the [X/Fe]$-$[Fe/H] space depends on both the energy and on the progenitor mass of the Pop~III SNe that enriched their birth clouds (see also \citealt{Vanni23}). However, we can notice that the descendants of low-energy Pop~III SNe (faint, cc) cover a wide range in both [Fe/H] and [C/Fe], $\rm -7.5\lesssim[Fe/H]\lesssim -1$, and, $\rm -1\lesssim[C/Fe]\lesssim 6$, and they are characterized by $\rm [Zn/Fe]\lesssim 0$. On the other hand, Pop~III descendants which are predominantly enriched by energetic SNe (high and hypernovae) exhibit a narrower range in $\rm -1\lesssim[C/Fe]\lesssim 3$ and $\rm -4<[Fe/H]<0$. Note that only stars imprinted by high energy SNe and hypernovae can reach super-solar [Zn/Fe] values, up to $\approx+1$\,dex.
Finally, in Fig.\ref{fig3}  {(bottom row)} we also show the regions covered by stars enriched by {\it{normal}} Pop~II/I SNe  {assuming different set of stellar yields}, which have $\rm [C/Fe]\lesssim1$ and they can reach high values of $\rm [Zn/Fe]\gtrsim 0$, comparable with those obtained with hypernovae enrichment. 

From comparing our model results with observational data for Boötes~I (Fig.~\ref{fig3}) it is evident that {\it none of the individual models} which assume a single Pop~III SNe type (only faint/cc/high/hyper) is able to reproduce, at the same time, the observed distributions of [C/Fe] and [Zn/Fe]. In fact, assuming that all pristine SNe evolve as low-energy SNe allows to reproduce the high values of [C/Fe] observed in Boötes~I stars but not the high [Zn/Fe] values. On the other hand, energetic SNe can reproduce the measured values of [Zn/Fe], but fail to reproduce the observed [C/Fe] stellar ratios.
Finally, even taking into account the contribution of Pop~II SNe and using different set of stellar yields/EDFs, it is not possible to reproduce the data, as well as assuming a single type of Pop~III SNe.
In conclusion, to reproduce the variety of [C/Fe] and [Zn/Fe] measured in Boötes~I stars with different [Fe/H] we need the contribution of Pop~III SNe with different energies.

 {Thus, we adopt as our \textit{\vl fiducial model\vd} the one that minimizes the reduced chi-squared: the \vl all types\vd\ EDF (see Fig.\ref{edf_teo}),  paired with the LC Pop~II stellar yields.} \\

\label{testing}



\begin{figure*}
\centering
	\includegraphics[width=0.65\textwidth]{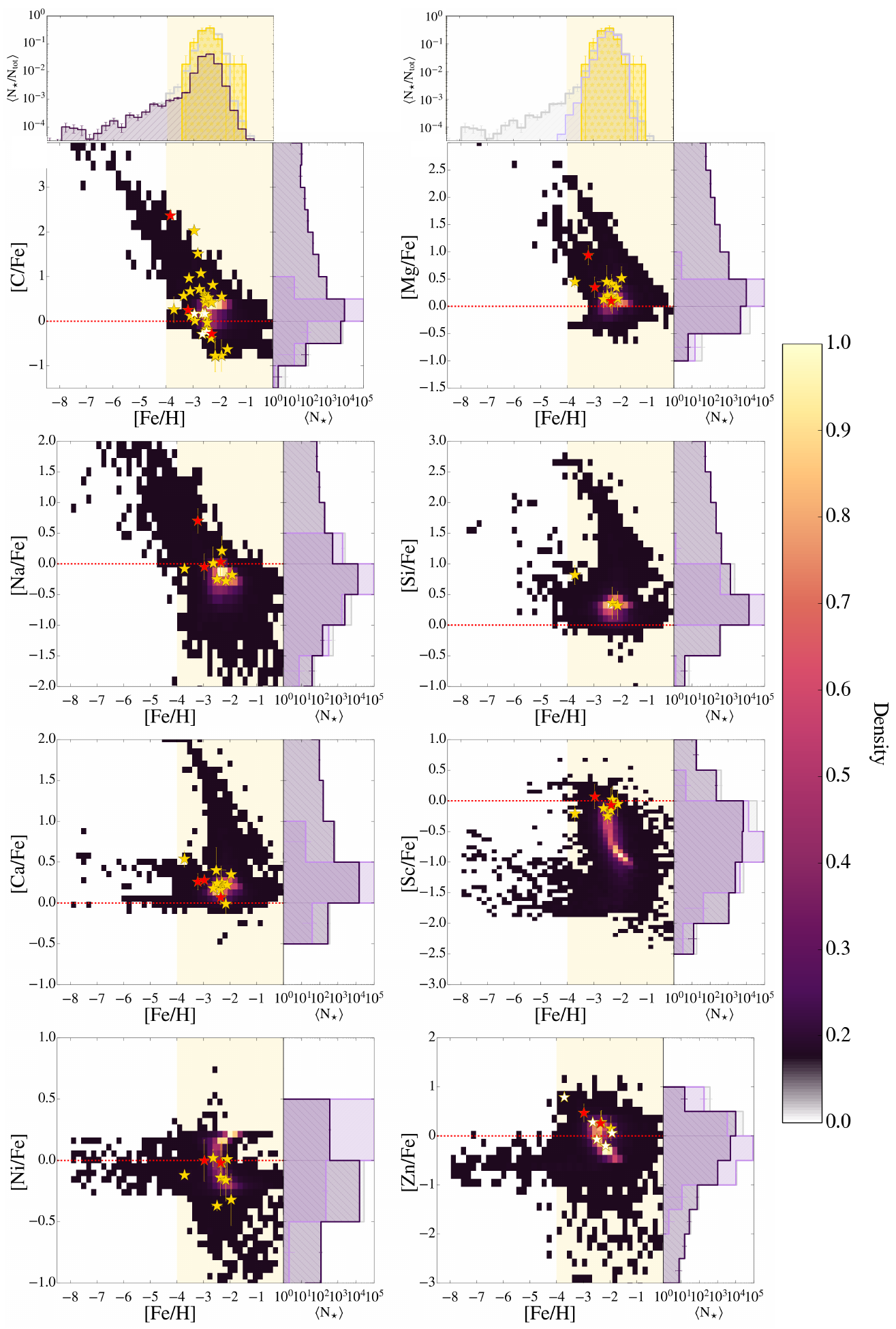}
   \caption{Density maps, [X/Fe] $-$[Fe/H],  of the simulated stellar populations at present day, $z=0$. Yellow star symbols are observed chemical abundances of Boötes~I stars (SAGA Database \cite{SAGA}) {, white symbols indicate stars for which [X/Fe] represents an upper limit, while red symbols denote potential candidates for Pop~III descendants.}. Top marginal histograms show the MDFs of:observed Boötes~I stars (yellow), all stars (light grey), stars enriched at level higher than $75\%$ by Pop~III stars (dark purple, top left) and stars enriched at level higher than $75\%$ by Pop~II stars (light purple, top right). Right marginal plots show the [X/Fe] abundance ratio distributions for:
   all simulated stars (light grey), and  predominantly ($>75\%$) Pop~III/Pop~II enriched stars (dark/light purple). Shaded yellow area highlights $\rm [Fe/H]>-4$. 
   }
   \label{xratio1}
\end{figure*}

\section{Early chemical evolution: predictions}

In Fig.~\ref{xratio1}, we show  the predicted [X/Fe] versus [Fe/H] density maps for the present-day stellar populations in Boötes~I predicted by our \textit{fiducial model} (\vl all types\vd\ in Fig.~\ref{edf_teo} top panel).
 
The first thing to note is that all the simulated [X/Fe] density maps effectively reproduce the observed data,  {with the exception of [Sc/Fe], where there is a discrepancy of approximately one dex between the observations and simulations. This discrepancy in [Sc/Fe] is a known issue in nucleosynthesis modeling and has been highlighted in previous studies (e.g., \cite{kobayashi20})}. Indeed,  in all other cases the peaks of the simulated density distributions for each chemical element perfectly match the corresponding observed distributions.  It is worth mentioning that the  simulated density maps include many more stars than those observed, and cover a broader range in the [X/Fe]-[Fe/H] space. Our model predicts regions of low density ($< 20\%$ of the density peak), representing rare stellar populations, not yet observed in UFDs, that are particularly intriguing for investigating the signatures of the first stars. \\
In the top marginal plots of Fig.~\ref{xratio1} we compare the normalized MDFs of Boötes~I observed and simulated, including the predicted MDFs for stars predominantly enriched by Pop~III ($>75\%$ of their metals, top right) and by Pop~II stars (top left). In the range where data are available, $\rm [Fe/H]>-4$, the simulated MDF is dominated by stars predominantly enriched by normal Pop~II SNe. Indeed, Pop~II enriched stars represent $>90\%$ of the total number of stars in these [Fe/H] range, while stars mainly imprinted by Pop~III SNe are extremely rare at $\rm [Fe/H]>-4$, only counting for $\approx 1\%$ of the total population.
 {In the low-metallicity tail of the MDF ($\rm [Fe/H] < -4$), more than $80\%$ of the stars are predominantly imprinted by Pop~III SNe. These Pop~III descendants can be further distinguished from Pop~II-enriched stars in this regime by their unique [X/Fe] abundance ratios, marking them as {\it outliers} in the chemical abundance space.}

\subsection{Key abundance ratios to unveil Pop~III descendants }
Now we investigate if there are regions in the [X/Fe]~$-$~[Fe/H] space where it is possible to \textit{uniquely} identify the fingerprints of Pop~III SNe.
To this aim we derive the X-Distribution Function (XDF), i.e. the number of stars in bins of [X/Fe] averaged over different runs, $\langle \rm N_{\star}\rangle$.

In the right marginal plots of Fig.~\ref{xratio1}, we show the comparison between the total XDF (grey), and the XDFs of stars predominantly (>75\% of their metals) enriched by Pop~III (dark purple) or normal Pop~II/I stars (light purple).

For all chemical abundance ratios, the peak of the XDF corresponds to the density peak in star formation and thus it is dominated by stellar populations mainly enriched by normal Pop~II/I stars as already explained \citep[see also][]{Rossi+21}. 
In the low density regions, however, the fingerprints of the first SNe arise, i.e. the XDFs are typically dominated by Pop~III enriched stars. 
In conclusion, stars characterized by $\rm [C/Fe] \gtrsim +1$ or $\rm [C/Fe] \lesssim 0$, $\rm [Mg/Fe] \gtrsim +0.5$, $\rm [Na/Fe] \gtrsim +0.5$, $\rm [Si/Fe] \gtrsim +1$, $\rm [Ca/Fe] \gtrsim +0.5$, $\rm [Ni/Fe] \gtrsim +0.5$, or $\rm [Zn/Fe] \lesssim -1$ have a high probability ($> 80\%$) to be Pop~III descendants. 

\begin{figure*}
	\centering
	\includegraphics[width=0.65\textwidth]{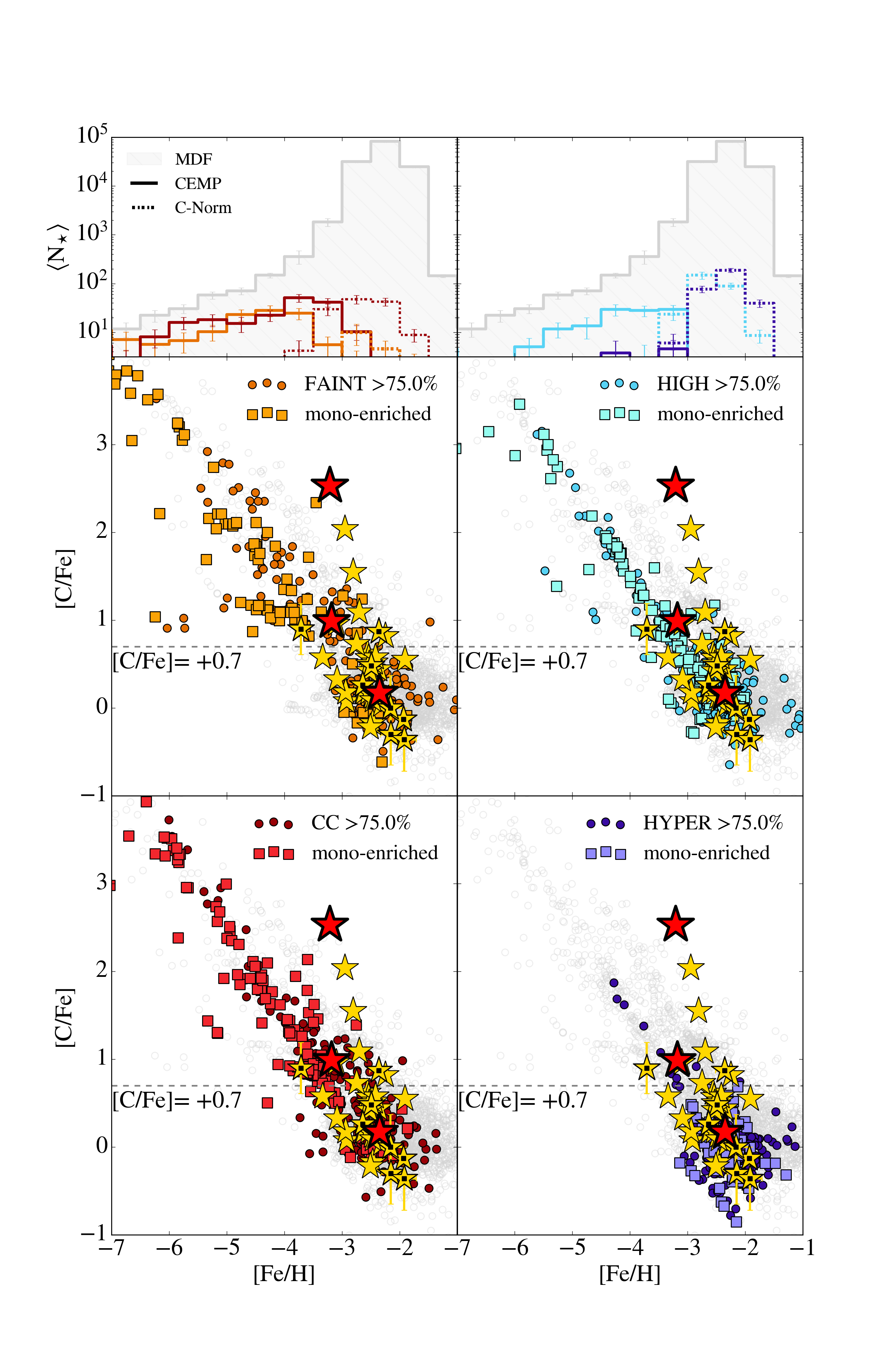}
   \caption{ The [C/Fe] vs [Fe/H] diagram of predicted Boötes~I stellar populations mainly enriched (>75\%) by Pop~III SNe with different energy: low-energy (first column), high-energy (second column). Top marginal plots show the average number of stars, $\rm \langle N_{\star} \rangle$, as a function of [Fe/H] for CEMP (solid) and C-normal (dashed) stars enriched at level higher than $75\%$ by different Pop~III SNe types, as well as the  overall MDF in Boötes~I. Yellow stars represent the Boötes~I data, while double marked points identify stars for which abundance pattern is available in the literature.  Red stars represent candidates for Pop~III descendants.}
   \label{C_fe_different_energies}
\end{figure*}


\subsection{The imprints of different Pop~III SNe types}
\label{imprint of different Pop III SNe}
According to our model predictions, a large fraction of stars with  $\rm [C/Fe] \gtrsim + 1$ (at $\rm [Fe/H]\lesssim -3.5)$ or $\rm [C/Fe] \lesssim 0$ (at $\rm[Fe/H] \gtrsim -3$) are likely to be Pop~III descendants (Fig.~\ref{xratio1}).
\noindent Now we go deeper into understanding the origin of these Pop~III descendants, exploring the {\it type} of SNe that enriched their birth cloud. In Fig.~\ref{C_fe_different_energies} we present the [C/Fe]-[Fe/H] diagram for Boötes~I stellar populations predominantly enriched by different Pop~III SNe types.
As we can see, Pop~III descendants can cover a very broad range of both $-7 < \rm [Fe/H] < -1 $ and $-1 \lesssim \rm[C/Fe] \lesssim 4$. However, the probability to find them strongly depends on their location in the [C/Fe]-[Fe/H] space, being maximum for $\rm [Fe/H]\lesssim -3.5$ and $\rm [C/Fe]\gtrsim +1$; or when $\rm [Fe/H]\gtrsim -3.5$ and $\rm [C/Fe]\lesssim 0$. This is also illustrated in Fig.~\ref{prob_pop} (Appendix \ref{appendix1}), where we show the absolute carbon abundance, $\rm A(C)$, with respect to $\rm[Fe/H]$.\\

At $\rm [Fe/H]\lesssim -3.5$, around $90\%$ of the stars in Boötes I are predicted to be CEMP stars,  {in which the carbon enhancement is driven by Pop~III SNe}. 
By inspecting the top marginal plots of Fig.\ref{C_fe_different_energies}, where the average number of stars, $\langle \rm N_{\star} \rangle$, predominantly imprinted by Pop~III SNe with different energies is shown, it is evident that the enrichment of CEMP stars can arise from different Pop~III sources: faint, cc and high energy SNe. Thus, we can conclude that the progenitors of Pop~III enriched CEMP stars in Boötes~I can be both low-energy (faint, cc) and high-energy Pop~III SNe. 

Conversely, direct descendants of hypernovae are uncommon within this range, $\rm [Fe/H]\lesssim -3.5$, and they mainly have [C/Fe]$<0.7$.\\

\noindent As the [Fe/H] increases, the population of CEMP stars enriched by low-energy Pop~III SNe, decreases; while the population of C-normal stars, enriched by energetic (high-energy and hyper) Pop~III SNe, increases (see right marginal plot of Fig.\ref{C_fe_different_energies}). Note that, towards high [Fe/H] values, the probability to find 
descendants of Pop~III stars declines due to the chemical pollution of subsequent Pop~II stellar populations. Therefore, we expect that the majority of stars with $\rm [Fe/H] \gtrsim -3.5$  have originated from gas that has been enriched by multiple generations of Pop~II stars (see Appendix~\ref{appendix1}). 

Nevertheless, among C-normal stars with $\rm[C/Fe]\lesssim +0.7$ we expect to find the imprints of Pop~III SNe of different energies, while the imprint of hypernovae is dominant at  $\rm[C/Fe]\lesssim -0.5$, accounting for $6\%$ of the total stars in this range.
\\

\subsection{Does Boötes~I host stars mono-enriched by Pop~III SNe?} 
Here, we define \vl mono-enriched\vd\: stars as those that have originated from gas enriched 
by {\it exactly one} Pop~III supernova. 
Fig.~\ref{C_fe_different_energies} shows that in Boötes~I we can potentially find mono-enriched stars which cover wide ranges in [Fe/H] and [C/Fe], and are enriched by different types of Pop~III SNe. In line to what we discussed already, for $\rm [C/Fe] > +0.7$ and $\rm [Fe/H]\lesssim -3.5$ we find predominantly mono-enriched stars imprinted  by faint, cc, and high energy SNe. Mono-enriched CEMP stars are rare $\sim 5\%$ with respect to the total of CEMP stars in Boötes~I, and  $\sim 5\%$ with respect to CEMP stars predominantly enriched by Pop~III.
On the other hand, mono-enriched stars by hypernovae are typically found at $\rm [C/Fe] < +0.7$ and $\rm  [Fe/H]\gtrsim -3.5$. Their fraction is maximum for $\rm [C/Fe]< -0.5$ representing $10\%$ with respect to the total number of stars in this range.


\section{Progenitors of individual Boötes~I stars}

We can now use the predictions of our chemical evolution model to uncover hidden Pop~III descendants in Boötes~I. To this end, we analyzed the measured abundance patterns of Boötes~I stars available in the literature, for which at least five chemical abundance ratios [X/Fe] have been measured.

To identify the progenitors of Boötes~I stars, we compared the abundance patterns of simulated stars that survive until today with those of observed stars. { Firstly we pre-selected models that have [Fe/H] consistent with the observed values, i.e., within the error bars.
Then, among all models reproducing [Fe/H], we found those that best fit all the measured chemical elements by minimizing the reduced chi-squared, $\chi^2$, between the simulated and observed [X/Fe] ratios:}
\begin{equation}
\chi^2 =  \sum_{i=1}^{N_{\text{obs}}} 
    \frac{\left( [\text{X/Fe}]_{\text{obs}, i} - [\text{X/Fe}]_{\text{sim}, i} \right)^2}
    {\sigma_{[\text{X/Fe}]\text{obs},i}^2 + \sigma_{[\text{X/Fe}]\text{sim},i}^2}
\end{equation}
 {where $
\rm [X/Fe]_{\text{obs}, i} $ represents the observed abundance with its corresponding uncertainty $\sigma_{ \rm [X/Fe]_{\text{obs}, i}}$, while$ \rm [X/Fe]_{\text{sim}, i}$  denotes the simulated abundance for the same element, with an assumed uncertainty of $\sigma_{\rm[X/Fe]_{\text{sim}, i}} = 0.2$.} Finally, after selecting the best fits, we retrieve from the simulation the properties of the SNe responsible for enriching the birth cloud from which these stars originated.
Note that the [C/Fe] ratio of the observed stars in Boötes~I have been corrected for the effect of evolutionary status by exploiting the online tool\footnote{http://vplacco.pythonanywhere.com/} presented in \cite{placco14}, while the abundance ratios are not corrected for 3D and Non-Local Thermodynamic Equilibrium (NLTE) effects. \\

\noindent We categorized Boötes~I stars into the following groups, according to their  {possible} progenitors: i) \textit{Hidden Pop~III descendants}, long-lived stars consistent with $100\%$ enrichment by Pop~III SNe. 
ii) \textit{Ambiguous stars}, whose abundance patterns are consistent with a predominant enrichment from both Pop~III  and/or Pop~II stars, making it challenging to distinguish between these two scenarios; and iii) \textit{Pop~II descendants}, which are primarily enriched by the normal Pop~II stellar population. 
Below, we will concentrate on the truly/ambiguous Pop~III descendants, i.e., groups i) and ii), while in Appendix~\ref{appendix2} we discuss group iii).

\subsection{Hidden Pop~III descendants}
\label{popiii_desc_sec}
Three stars in Boötes I have abundance patterns consistent with a $100\%$ enrichment from Pop~III SNe,  {therefore making them good candidates for being Pop~III descendants.}. When we refer to a $100\%$ enrichment by Pop~III SNe, we are indicating that the star formed in an environment enriched solely by Pop~III SNe. However, this does not imply that the star has been enriched by a single supernova, i.e. that the star is a mono-enriched Pop~III descendant. For this reason, in the following, we will differentiate between multi- and mono-enriched Pop~III descendants. 
The comparison between the measured abundance patterns and the best-fits ones are shown in Fig.\ref{abundace_pattern}.  {The figure also displays the two models with $\chi^2$ closest to the minimum, which in all cases also have 100\% Pop~III enrichment. Additionally the best fit for 100\% Pop~II enrichment is shown, but this has significantly higher $\chi^2$ in all three cases.}

\subsubsection{Mono-enriched Pop~III descendants}
In Boötes~I we uncover one  {potential mono-enriched Pop~III descendant }, i.e. a star imprinted by a single Pop~III SNe:\\

\begin{figure*}
	\centering
 \includegraphics[width=\textwidth]{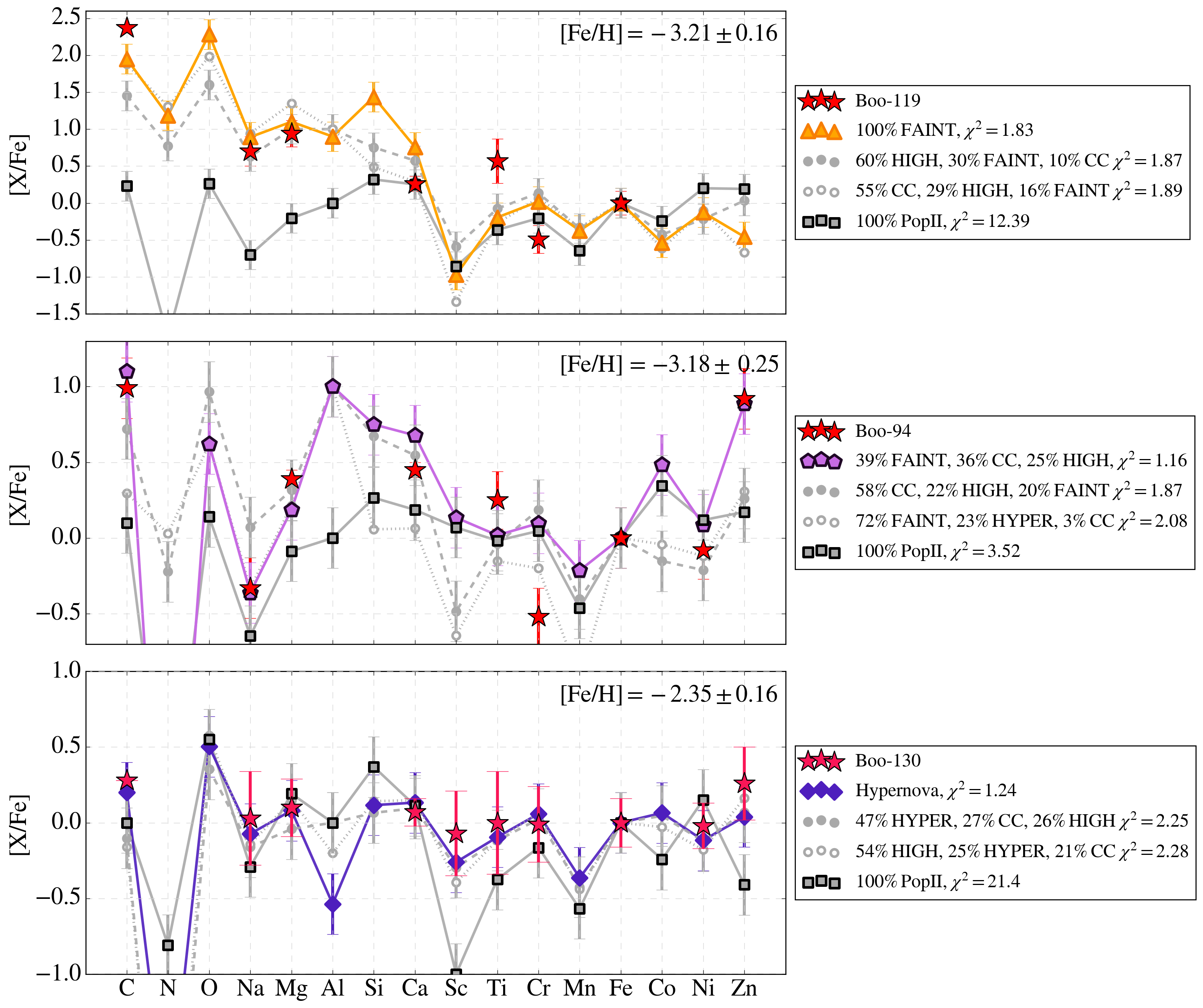}
   \caption{
   { Measured abundance patterns of three Boötes~I stars classified as {\it hidden Pop~III descendants} (red star symbols) compared to our best fits (colored lines). From top to bottom we show the data of the CEMP-no Boo-119 star \citep{lai11,gilmore13}; the CEMP-no star Boo-94 \citep{ishigaki14,gilmore13}; and the C-normal star Boo-130 \citep{ishigaki14}. The second (filled circles, dashed lines) and third (open circles, dotted lines) best fits are also shown in gray along with the Pop~II enrichment scenario with the lowest $\chi^{2}$ (squares, solid lines)}.
   }
   \label{abundace_pattern}
\end{figure*}

     {Boo-130:} The best fit indicates that Boo-130 is mono-enriched by a Pop~III hypernova with $m_{\star} = 64 \msun$ and $\rm E_{SN} = 5 \times 10^{51}$ erg (Fig.\ref{abundace_pattern}). Note that in this case, [Zn/Fe] is the key abundance ratio to discriminate between an enrichment driven by Pop~III and Pop~II stellar populations.  {Importantly, the models with $\chi^2$ values closest to the minimum also point to a $100\%$ Pop~III enrichment, in this case by multi-enrichment. This strengthens the hypothesis that Boo-130 is a strong candidate for being a Pop~III descendant.  While some degeneracy remains with other models featuring comparable  $\chi^2$-values, Boo-130 can be considered a descendant of Pop~III stars if we rely on the current model constraints}. \\

\subsubsection{Multi-enriched Pop~III descendants}

Among the three identified  {possible} Pop~III descendants, we found that two of them are {\it multi-enriched} by Pop~III SNe with different masses and/or energies:\\

     {Boo-119:} The measured abundance pattern of the CEMP ($\rm [C/Fe] \approx +2.5$) star Boo-119 (\citealt{lai11} for [C/Fe], \citealt{gilmore13} for rest) is consistent with $100\%$ {\it faint} Pop~III SNe enrichment (top panel of Fig.\ref{abundace_pattern}). However, according to our predictions, Boo-119 was enriched by {\it two} faint Pop~III SNe with similar progenitor masses: i)~$\rm E_{SN}= 0.3 \times 10^{51}\: erg$ and $m_{\star}= 43 \msun$ (50\%); and ii)~$\rm E_{SN}= 0.3 \times 10^{51}\: erg$ and $m_{\star}= 55 \msun$ (50\%). As we can see, our predicted abundance pattern nicely reproduces that observed except for [Ti/Fe], which is commonly underestimated in models compared to observations.  \\
    Conversely, the best fit among all the abundance patterns solely driven by a Pop~II SN enrichment, fails to reproduce the observed [C/Fe] value, which is lower by $> 1$ dex,  {as well as the [Mg/Fe] and [Na/Fe] ratios, which exhibit a discrepancy of $\gtrsim 0.8$ dex. Note that the other models with  $\chi^2$-values closest to the minimum also predict an enrichment driven by $100\%$ Pop~III SNe, albeit with different contributions from different Pop~III SNe types (Fig.\ref{abundace_pattern}).} \\
  
     {Boo-94:} The measured abundance pattern of the Boo-94 star is consistent with $100\%$ enrichment by  {{\it three}}  {Pop~III SNe with different energies ($\sim 39\%$ by faint, $\sim 36\%$ by cc, $\sim 25\%$ by high-energy) and masses: with $\rm E_{SN}= [0.3, 1.2, 3] \times 10^{51}\: erg$ and masses respectively equal to $m_{\star}= [24, 30 , 35] \msun$. Also for this star, the predicted abundance pattern of a  $100\%$ Pop~II enrichment fails to accurately reproduce the observed abundances , in particular in regards to in [C/Fe], [Na/Fe] and [Zn/Fe], see Fig.\ref{abundace_pattern}.} \\

 \subsection{Ambiguous stars}
\label{ambiguos_sec}

\begin{figure}
	\centering
\includegraphics[width=\columnwidth]{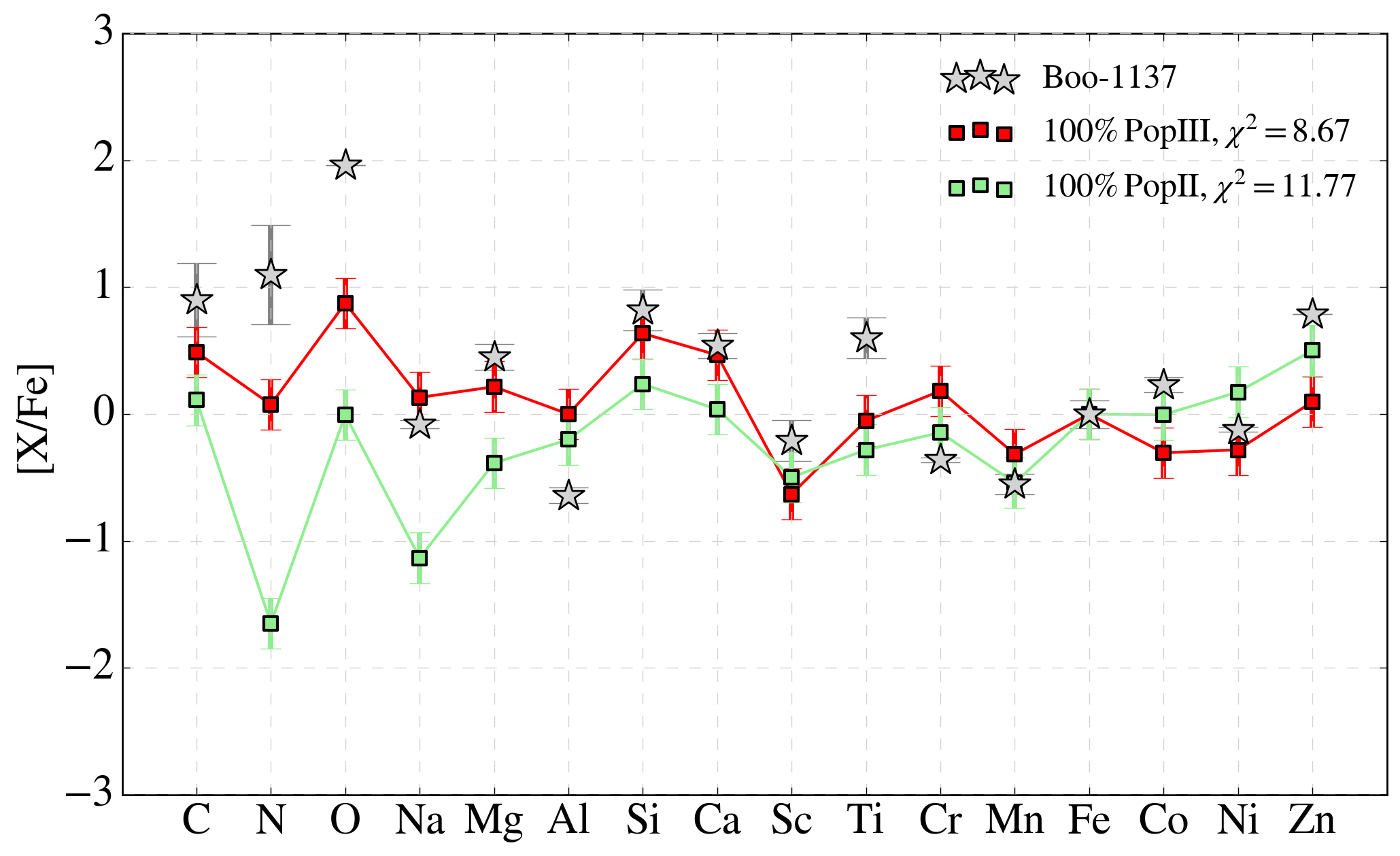}
 \caption{Comparison between the measured abundance pattern of the star Boo-1137 classified as {\it ambigous} (grey star symbols, \citealt{norris10}) and our best-fit models (squares) for an enrichment driven by Pop~II (green) and by Pop~III stellar (red) populations.}
   \label{ambiguous}

\end{figure}

Within the analyzed Boötes~I stars, we have identified a subset that we now classify as \vl ambiguous\vd. An example of this is displayed in Fig.\ref{ambiguous},
where we show the comparison between the measured abundance pattern of Boo-1137 (\citealt{norris10}) and the simulated one that produce the best fit. For this star, the minimum $\chi^2 \approx 9$ is obtained with an enrichment mainly driven by Pop~III SNe.
However, when plotting the predicted abundance pattern for a 100\% Pop~II enrichment  with a $\chi^2$-value very close to the minimum ($\chi^2 \approx 12$), we observe that these two abundance patterns appear strikingly similar. Specifically, the two predicted abundance patterns are almost indistinguishable in the  C, Al, Si, Ca, Sc, Ti, Cr, Mn, Co, Ni, Zn abundances, while they exhibit differences of $> 1\,\rm dex$ in  N, O, and Na ,  {for which no measurements are available}.
\noindent It is worth noting that this degeneracy could be resolved with the measurement of [N/Fe]. However, nitrogen is both challenging to measure accurately and the surface abundance strongly depends on the stellar evolutionary stage, due to internal stellar mixing processes \citep[e.g.][]{gratton00,placco14}.

\rm \textit{Why an enrichment driven by 100\% Pop~III is so similar to 100\% Pop~II SNe?}
The reason lies in the number of Pop~III SNe progenitors contributing to the ISM pollution. Indeed, upon retrieving the progenitors, we found that this abundance pattern is the result of enrichment from multiple Pop~III SNe (>10), with different masses and energies.  Consequently, the distinctive features of individual Pop~III SNe are lost resulting in an abundance pattern very similar to that driven by Pop~II stars (see Fig.\ref{ambiguous}).

\section{Probing Pop~III Supernova properties}

\begin{figure}
	\centering
\includegraphics[width=\columnwidth]{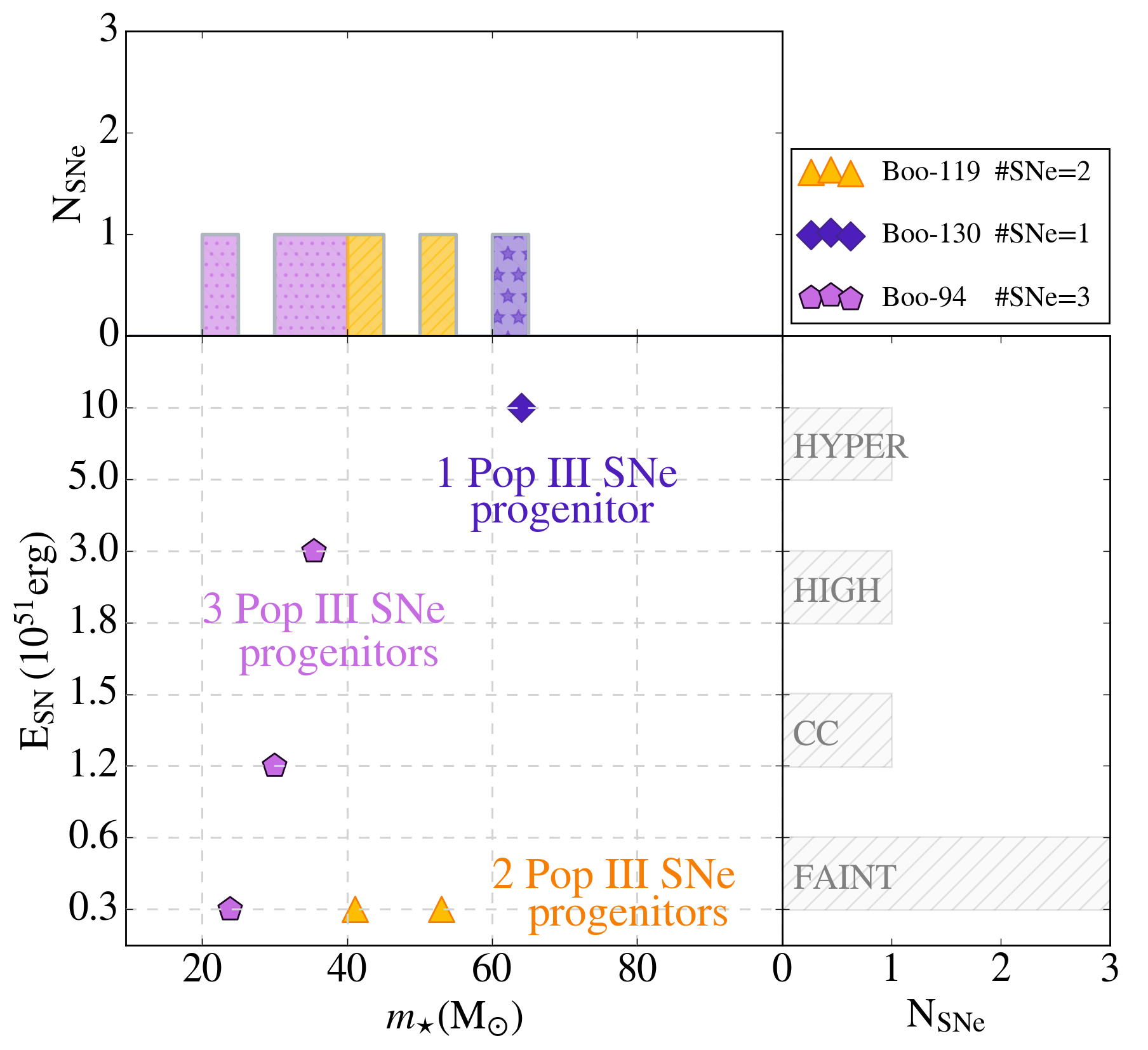}
 \caption{The energy-mass ($\rm E_{SN}$-$m_\star$) distribution of Pop~III SNe progenitors of 
 Boötes I stars. The two multi-enriched first stars descendants Boo-119 (orange) and Boo-94 (purple); and a mono-enriched Pop~III descendant Boo-130 (blue). Marginal plots show the number of SNe ($\rm N_{SNe}$) as a function of $m_\star$ (top) and $\rm E_{SN}$ (right).  }
   \label{new_approach}

\end{figure}

\noindent In Fig.~\ref{new_approach}, we show the energies ($\rm E_{SN}$) and masses ($m_{\star}$) of the Pop~III SNe progenitors for our sample of Pop~III descendants (Boo-119, Boo-130, and Boo-94; see Fig.\ref{abundace_pattern}).
As we can see in the marginal plots, our results probe the existence of Pop~III SNe in the mass range $m_{\star} = [20-65] \msun$ and show the coexistence in Boötes~I of Pop~III descendants with different energies, from faint to hypernovae.

 {Our analysis strongly suggests that these stars are compelling candidates for being genuine Pop~III descendants. However, as illustrated in Fig.\ref{abundace_pattern}, a clear degeneracy remains: different combinations of Pop~III SNe types, with varying progenitor masses and explosion energies, can produce equally good fits to the observed abundance patterns. While chi-square minimization identifies the best-fitting models, it is important to note that multiple solutions exist within a narrow range of $\chi^2$ values. Here the aim is to introduce a method to infer the physical properties of the Pop~III progenitors, such as their masses and explosion energies, which shaped the chemical abundances observed in these stars. Naturally, obtaining more robust constraints will require additional measurements of chemical elements, which could help break the degeneracies and better differentiate between competing scenarios. Nonetheless, with the currently available data, we provide the most reliable estimates possible.}
\subsection{Metal Retention in Boötes~I}
\begin{figure}
	\centering
\includegraphics[width=0.8\columnwidth]{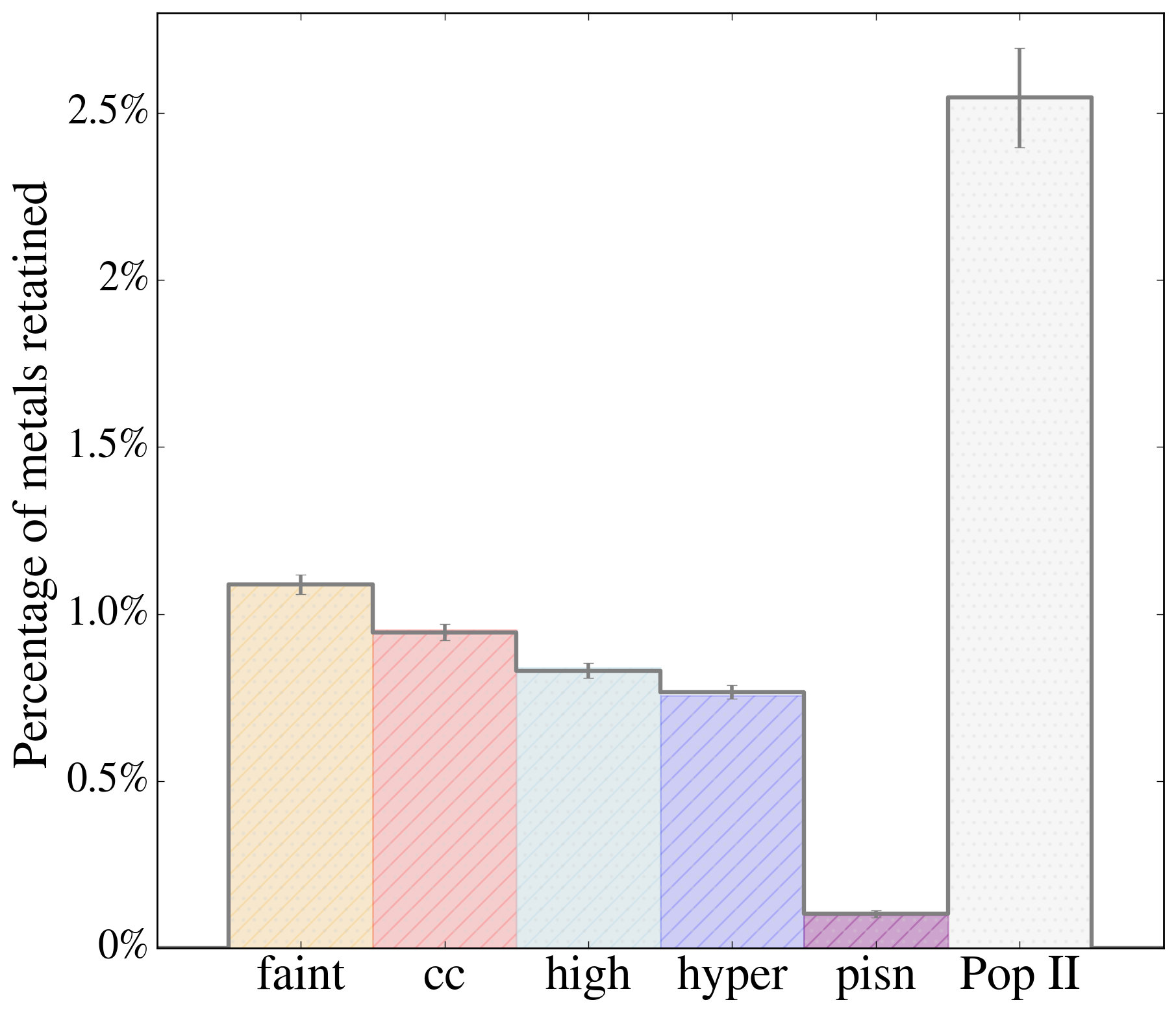}
 \caption{The fraction of metals retained in the potential well of Boötes~I galaxy after the explosion of different SNe type: faint (orange), cc (red), high-energy (light blue), hypernovae(blue), PISN (purple), Pop~II SNe (grey).}
   \label{metals}

\end{figure}

\noindent According to our model, the three Pop~III SNe descendants have been enriched by a total of  {three} faint, one cc, one high-energy SN and one hypernovae, i.e. four out of six  Pop~III progenitors are low-energy SNe.
Despite the limited statistics, our findings seem to suggest that Boötes~I preferentially retains the chemical elements released by low-energy Pop~III SNe. 
To further explore this aspect, in Fig.~\ref{metals} we evaluate the percentage of metals, coming from different SNe types, that are locked in stars surviving until $z=0$. Given that Boötes I is currently a gas-free galaxy, this percentage also reflects the metals retained by its gravitational potential well throughout its evolution.\\
\noindent The first thing to note is that the fraction of metals retained by stars is always very small, $\lesssim 2.5\%$, the maximum corresponding to the case of Pop~II SNe. The percentage is even lower for Pop~III SNe, less than $\lesssim 1.1\%$ of metals retained. 
Moreover, for Pop~III SNe, there is clear trend with explosion energy: the fraction of metals retained is greater $\gtrsim 1\%$ for low-energy SNe (faint, cc) but decreases as the energy increases. 
\noindent Finally, in Fig.~\ref{metals} we also show the fraction of metals retained by Boötes I when PISN explode. As observed, the fraction is $< 0.1\%$. This is attributed to two distinct effects: firstly, the random sampling of the IMF plays a fundamental role.  Indeed, when the first stars form, the star-formation rate is low (less than $10^{-2} \msun \rm yr^{-1}$, see \citealt{Rossi+21}), making the formation of stars in the PISN mass range, i.e., $[140-260]\msun$, less probable.
Secondly, when PISNe explode, they are so energetic that they blow out all the gas within the galaxy. Only in the case of low-mass PISNe ($m_{\star} \sim 140 \msun$) evolving, a fraction of the metals injected into the ISM can be retained by Boötes~I, however the probability to form them is very low. We can conclude that the probability to find the imprint of PISN in UFDs is extremely low due to their shallow potential wells.

\section{Discussion}
The approach of comparing observed and simulated abundance patterns to determine the properties of progenitors of metal-poor stars is widely used. However, this approach typically involves fitting the observed abundance patterns with the yields from a single Pop~III SN \citep[e.g.][]{Heger+woosley10, ishigaki18} or a few \citep[e.g.][]{Hartwig+18, Welsh23} Pop~III progenitors.
Finding stars enriched by a single Pop~III SNe is exciting, but we have demonstrated that such stars are expected to be extremely rare even in ancient and metal-poor UFDs. The majority of Pop~III descendants are expected to be enriched by multiple progenitors, with different masses and energy, each contributing to the observed abundance patterns with different percentage (see Sec.\ref{popiii_desc_sec}). Therefore, it is crucial to interpret these data using {\it self-consistent} chemical evolution models that account for the different sources of enrichment. We show that identifying Pop~III descendants is possible, though challenging, and that some may be hidden in existing literature data.

\noindent Among the 54 Boötes~I stars with measured chemical abundance ratios, three are consistent with being direct Pop~III descendants. Notably, two of these three stars are multi-enriched by Pop~III SNe with different energies and masses (see Sec.\ref{popiii_desc_sec}). On the other hand, our results show that if the gas has been enriched by more than 10 Pop~III SNe, the resulting abundance pattern does not display the distinctive chemical features of Pop~III descendants. Instead, it becomes indistinguishable from that produced by Pop~II stars (see Sec.\ref{ambiguos_sec}). Boo-1137 is a perfect example of this case (see Fig.\ref{ambiguous}), where we cannot discriminate between enrichment driven by a Pop~III or Pop~II/I stellar population. This star was also analyzed by \cite{Welsh23}, who found that Boo-1137 had been enriched by a single Pop~III progenitor with \(m_{\star} = (18-27) \msun\). It is important to note that to achieve this result, the authors used only five chemical elements (C, O, Al, Si, Fe), while for Boo-1137, there are 11 other measured chemical elements in the literature that we used to determine the best-fit model. We can speculate that the different interpretation of the origin of this Boo-1137 is due to the number of chemical abundance ratios used in the analysis.  This discrepancy emphasizes the importance of employing a wide range of chemical elements (> 5) to accurately interpret data and to give constraints on the initial metallicity, mass and explosion energy of the progenitor stars imprinting the birth environment. \\

It is worth noting that all our results are based on our fiducial model, which adopts a flat EDF for PopIII stars and the PopII stellar yields from \citet{Limongi+18}. To test the robustness of our interpretation, we investigated whether the abundance patterns of our PopIII candidates could also be reproduced using alternative PopII enrichment scenarios. Specifically, we employed the yield tables from \citet{NK13} at a fixed metallicity of $\log(Z/Z_{\odot}) = -3$, considering both standard core-collapse SNe and hypernovae. For each of our three Pop~III-only candidates (Boo-119, Boo-130, and Boo-94), we computed the expected abundance patterns assuming enrichment from either a single Pop~II SN or a combination of two Pop~II progenitors with different masses. As shown in Fig.\ref{PopIIdifferentyields} of Appendix \ref{appendix3}, adopting these alternative PopII sets of yields, we obtain best-fit models with higher chi-squared values compared to the PopIII-only enrichment scenarios, i.e., our fiducial model. This supports the interpretation that these stars are robust candidates for being genuine Pop~III descendants.

For direct Pop~III descendants we infer the mass of their progenitors that is in mass range $ m_{\star}=[20-65] \msun$. These results are partially in agreement with the results of \cite{ishigaki18}. The authors found that the abundance patterns of extremely metal-poor stars ($\rm [Fe/H]< -3$) are predominantly best-fitted by SNe yields with $m_{\star}< 40 \msun$ and that more than half of the stars are best-fitted by the $m_{\star} = 25 \msun $ hypernova $(E=10 \times 10^{51} \rm erg)$ models. Finally, they conclude that the masses of Pop~III stars responsible for the first metal enrichment are predominantly $m_{\star} < 40 \msun$. On the other hand, our results suggest that the masses of Pop~III stars responsible of the enrichment are higher, up to $ \approx 65\msun$. 
Note that \cite{ishigaki18} use different sets of stellar yields (see their Sec.2) and they have, for $m_{\star}> 25 \msun$, only two stellar model yields for stars with $m_{\star}= 40 \msun$ and  $m_{\star}= 100 \msun$ \citep[see Table~1 in][]{ishigaki18}. While we used the stellar yields of \cite{Heger+woosley10} that have more stellar models within this range. 

\noindent Finally, our results show that UFDs are strongly affected by SNe feedback processes (Fig.~\ref{metals}). In particular, we show that the chemical products of energetic PISN are completely lost by Bootes~I, which is hosted by a $\approx 10^7\msun$ mini-halo. This is in perfect agreement with simulations of Pop~III SNe in high-z low-mass minihalos \citep{bromm02,Ritter2012, Smith2015}. To quantitatively compare our predictions with other studies for nearby dwarf galaxies we evaluate the mass-loading factor\footnote{$\eta=\dot{M_{ej}}/\dot{M_{\star}}$, defined as the ratio between the rate of ejected mass and star formation rate.}, $\eta = 85 \pm 12$ that is in good agreement with the results of \cite{standford23} and with hydrodynamic simulations (e.g. \citealt{emerik19, pandya2021}).

\section{Conclusions}

To investigate the chemical signatures left by Pop~III SNe, we implement a theoretical chemical evolution model of Boötes~I \citep{Rossi+21, Rossi23}, the best studied UFD galaxy (see also \cite{romano14, vincenzo14, Alexander23}). Our model accounts for the incomplete sampling of the IMF in this poorly star-forming system. The chemical enrichment from carbon to zinc is followed by including SNe, Type Ia SNe (SN Ia) and AGB stars from both Pop~III stars and the subsequent Pop~II stellar populations. Our key results are:
\begin{itemize}
    \item To accurately reproduce the measured abundance ratios in Boötes~I stars (in particular [C/Fe] and [Zn/Fe]), it is necessary to include contribution of Pop~III SNe with {\it both low and high energies} (Fig.\ref{fig3}). Consequently, in our \vl fiducial model\vd\  we adopt a theoretical EDF which is distributed among different equiprobable energies (faint, cc, high-energy and hypernovae), see Fig.\ref{edf_teo}.
    
    \item In [X/Fe] vs [Fe/H] diagrams, true Pop~III descendants can be identified within low-density regions as {\it outliers}
     (see Fig.\ref{xratio1}). In particular, stars at $\rm [Fe/H] < -4$ with  $\rm [C/Fe] \gtrsim +1$ or $\rm [C/Fe] \lesssim 0$, $\rm [Mg/Fe] \gtrsim +0.5$, $\rm [Na/Fe] \gtrsim +0.5$, $\rm [Si/Fe] \gtrsim +1$, $\rm [Ca/Fe] \gtrsim +0.5$, $\rm [Ni/Fe] \gtrsim +0.5$ or $\rm [Zn/Fe] \lesssim -1$, have a higher probability to be Pop~III descendants.

    \item The [C/Fe] ratio is a {\it key abundance ratio} to unveil the descendants of Pop~III SNe. We show that 
    at $\rm [Fe/H]< -3$ we can find the imprint of Pop~III SNe with explosion energies $\rm E_{SN} = [0.3-3] \times 10^{51} \rm \: erg$, i.e. faint, cc, high-energy Pop~III SNe  among CEMP stars (Fig.\ref{C_fe_different_energies}). While the chemical signature of hypernovae arises at $\rm [Fe/H]>-3.5$ among C-normal stars with $\rm [C/Fe] < + 0.7$.
     
     \item Among literature data of Boötes I stars with $>5$ abundance ratios measured, we uncover {\it one mono-enriched and two multi-enriched Pop~III descendants} (Fig.\ref{abundace_pattern}). Specifically, our findings show that the CEMP star Boo-119 exhibits an abundance pattern consistent with enrichment by two faint Pop~III SNe;  Boo-94 formed in an environment polluted by three Pop~III SNe (one faint, one ccSN and one high-energy SN); and, notably, we identify a  mono-enriched Pop~III hypernova descendant, Boo-130; \\

     \item Retrieving properties of the progenitors of Pop~III descendants we found that they have masses in the range $[20-65]\msun$ and  {\it different energies} spanning from low to high energy (Fig.\ref{new_approach}).

      \item Examining the chemical evolution of Boötes~I, we found that in UFDs the fraction of metals retained by their gravitational potential is very low ($\lesssim 2.5 \%$) independent of the SNe type. 
      For low-energy Pop~III SNe the fraction is $\approx 1\%$, while it is even lower for more energetic Pop~III SNe (Fig.\ref{metals}). 

      \item Finding the chemical signatures of energetic PISNe is unlikely
      in systems like UFDs, as the fraction of metals retained by their potential wells is $\lesssim 0.2\%$ (Fig.\ref{metals}). 
      
\end{itemize} 

To delve deeper into unraveling the mysteries of Pop~III SNe, such as their energy and mass properties, it is crucial to identify their descendants and measure their chemical properties. 
Our analysis demonstrates that for understanding the chemical evolution of Boötes I it is vital to consider different Pop~III SNe types, whose direct descendants can be identified as {\it outliers} in different chemical abundance spaces. Overall, our study offers a robust framework for interpreting observational data in Boötes I, allowing for the identification of Pop~III descendants based on their chemical signatures. 
In conclusion, our model stands ready to analyze and interpret upcoming data like 4MOST/4DWARFS (\citealt{4dwarfs}), providing a comprehensive understanding of the properties of the first supernovae.

\section{Acknowledgments}.
\begin{acknowledgments}
This project has received funding from the European Research Council Executive Agency (ERCEA) under the European Union's Horizon Europe research and innovation program (acronym TREASURES, grant agreement No 101117455”). M.R., S.S., I.V., and I.K. acknowledge ERC support (grant agreement No. 804240). S.S. and I.V. also acknowledge support from the PRIN-MIUR2017, prot. n. 2017T4ARJ5. We thank Stefano Ciabattini for insightful discussions and for sharing his new findings.
\end{acknowledgments}

%




\appendix

\section{The probability to find Pop~III descendants}
\label{appendix1}
\label{imprint different energies}
The [C/Fe] ratio is commonly used as a diagnostic tool for identifying metal-poor stars enriched by Pop~III sources . Additionally, the A(C) value is employed to discern the origins of distinct CEMP stellar populations \citep{bonifacio15, Yoon19}: in \cite{Rossi23} we show that among CEMP-no stars, true Pop~III stars {\it descendants} appear for $ \rm A(C) \lesssim 6$ and $\rm [Fe/H] \lesssim -3.5$.\\

\begin{figure*}
	\centering
	\includegraphics[width=0.9\textwidth]{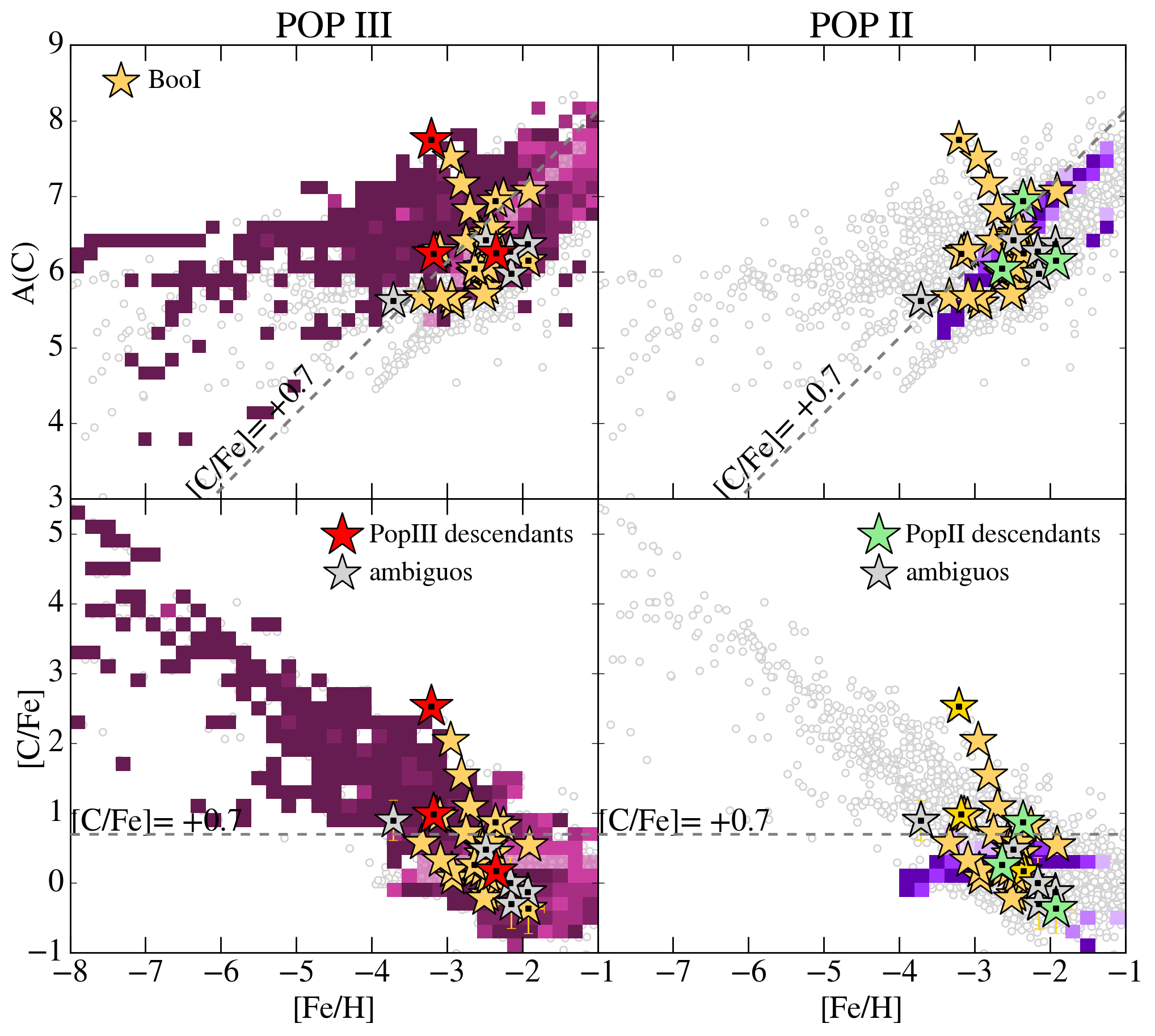}
   \caption{ The A(C)(top panel) and [C/Fe] (bottom panel) versus [Fe/H] diagram for Boötes~I stellar populations. The colored squares identify the probability to find, at fixed [Fe/H] and A(C) (of [C/Fe]), stars predominantly enriched by Pop~III or by Pop~II stars (columns). Star points represent the Boötes~I data where the ones with a square in their center identify the stars for which the abundance ratio of different chemical elements is available in the literature. The different color of Boötes I stars represent Pop~III descendants (red, Sec.\ref{popiii_desc_sec}), Pop~II descendants (green, Sec.\ref{popii_desc_sec}) and \vl ambiguous \vd stars (grey) for which it is not possible discriminate their origin, respectively (see Sec.\ref{ambiguos_sec}).}
   \label{prob_pop}
\end{figure*}

\noindent To unveil if there exists a region in which Pop~III descendants can be found,
we derived the probability that a given stellar population has been imprinted by Pop~III or Pop~II SNe for different [Fe/H] and [C/Fe] (or A(C)) values. 
\noindent  { Analyzing the A(C)-[Fe/H] diagram in Fig.~\ref{prob_pop}, a clear separation emerges between stars imprinted by Pop~III and Pop~II enrichment. Stars primarily enriched by Pop~III SNe are characterized by $4 \lesssim \rm A(C) \lesssim 8$ and are predominantly located at $\rm [Fe/H] < -3.5$. In contrast, stars primarily enriched by Pop~II SNe exhibit $\rm [Fe/H] > -3.5$ and none of them fall within the CEMP star category.Indeed, Pop~II SNe stellar yields adopted \citep{Limongi+18}, do not allow to form stellar populations enhanced in carbon, i.e. with $\rm [C/Fe] > +0.7$.}
\noindent Pop~III enriched stars are located in the low-branch at $ \rm A(C) \lesssim 6$ of the A(C)-[Fe/H] or equivalently in the bottom branch of the [C/Fe]-[Fe/H] diagram. The majority of Pop~III stars descendants with $\rm [Fe/H] < -4$ are CEMP-no stars and the probability to find them increases as [Fe/H] decreases and it is maximum for $ \rm A(C) \lesssim 6$ and $ \rm [Fe/H] \lesssim -3.5$. Furthermore, it is predicted that Pop~III descendants can be found among C-normal stars. In particular at $\rm [C/Fe] <0$ and $\rm -4 \lesssim [Fe/H] \lesssim -1$ the probability to find the chemical signatures of pristine SNe exceeds $ \gtrsim 50\%$. In conclusion, by exploiting both [C/Fe] and A(C) as diagnostics we can successfully predict promising descendants among both CEMP and C-normal stars: carbon-enhanced Pop~III stars descendants are predicted to have $ \rm A(C) \lesssim 6$ and $\rm [Fe/H] \lesssim -3.5$ while C-normal stars imprinted by Pop~III SNe have $\rm [C/Fe]< 0$ and $\rm [Fe/H] \gtrsim -3.5$. In other words 
 we predict that stars with these properties have the highest probability of being Pop~III descendants.\\

\vspace{2pt}
\section {Pop~II descendants}
\label{appendix2}
\label{popii_desc_sec}
\begin{figure}
	\centering
\includegraphics[width=0.5\columnwidth]{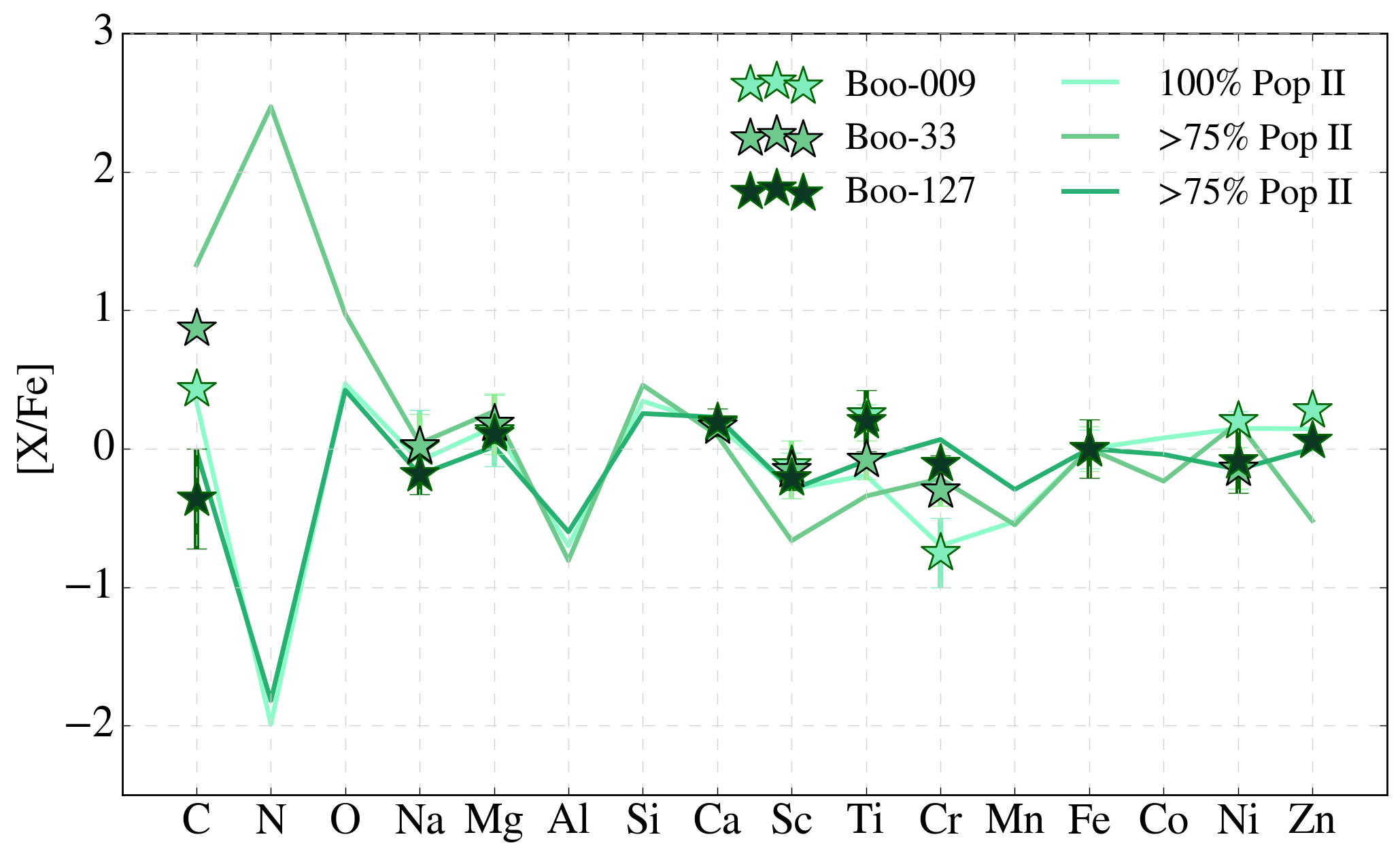}
 \caption{Comparison between the observed (star symbols) and the best-fit (lines) abundance pattern of Boo-009, B00-33, and Boo-127 \citep{gilmore13, ishigaki14}.}
   \label{PopII abund}

\end{figure}

\begin{itemize}
    \item  {Boo-009, Boo-33, Boo-127 -} We analysed the abundance pattern of Boo-009 (\citealt{ishigaki14}), Boo-33 (\citealt{gilmore13}) and Boo-127 (\citealt{ishigaki14}) (Fig.\ref{PopII abund}). The first thing to note is that the abundance patterns measured for these stars are very similar. Furthermore, they are in agreement with an enrichment predominantly driven by Pop~II SNe, accounting for $ 51\%$ in the best model for Boo-009, $82\%$ for Boo-33,  and  $85\%$ for Boo-127. While Boo-009 is a pure Pop~II enriched star that has been also enriched (at $49\%$) by Pop~II AGB, it is worth noting that both of Boo-33 and Boo-127 experienced some level of enrichment from Pop~III SNe hypernovae and core-collapse, which, however, did not leave particularly distinct signatures in their abundance patterns.
\end{itemize}

\section {Testing alternative Pop~II stellar yields}
\label{appendix3}
\begin{figure*}
	\centering
\includegraphics[width=0.8\textwidth]{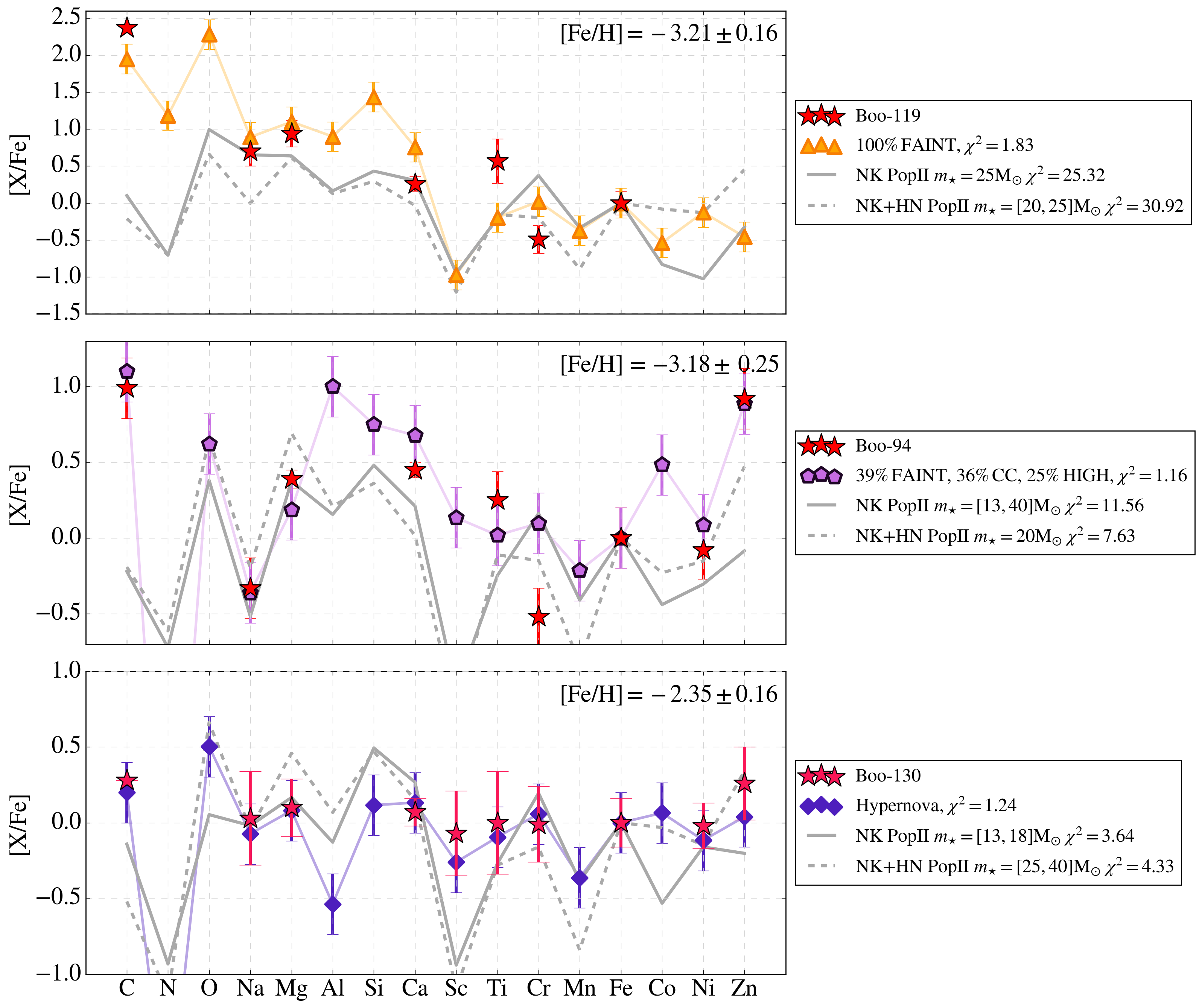}
 \caption{Comparison between observed abundance patterns of the three Pop~III candidate stars Boo-119, Boo-130, and Boo-94 (colored points) and synthetic abundance patterns predicted from Pop~II-only adopting the stellar yields from \citet{NK13} at $Z = -3$, including both standard core-collapse SN (NK, solid gray line)  and hypernovae (HN, dashed gray line). The panels show the model that yields the lowest $\chi^2$ value in each case, reported in the legend along with the corresponding progenitor mass or mass pair. In all three cases, Pop~II-only enrichment produces significantly worse fits compared to the Pop~III-only fiducial model.
}
   \label{PopIIdifferentyields}

\end{figure*}

In Fig.\ref{PopIIdifferentyields} we compare the observed chemical abundance patterns of our three Pop~III candidate stars (Boo-119, Boo-130, and Boo-94) with synthetic abundance patterns predicted from PopII-only enrichment scenarios. The models are computed using the stellar yield tables from \citet{NK13} at Z = -3, considering both standard core-collapse SNe and hypernovae. For each star, we test enrichment from either a single Pop~II progenitor or a combination of two with different initial masses. The resulting chi-squared values are shown for each scenario and compared to those obtained using the PopIII-only enrichment in our fiducial model. In all cases, the PopIII models provide significantly better fits, supporting their identification as genuine PopIII descendants.

\bibliography{hidden}{}
\bibliographystyle{aasjournal}



\end{document}